\documentclass[showpacs,showkeys,preprintnumbers,amsmath,amssymb,%
               twocolumn,byrevtex,a4paper,pra,aps,nofootinbib]{revtex4}
\usepackage[dvips]{graphicx}
\usepackage[english]{babel}
\makeatletter
\DeclareRobustCommand\abbr[1]{\let\last=\relax\@tfor\a@token:=#1\do%
  {{\sc\last}\let\last=\a@token}\futurelet\n@token\terminate@abbr}
\def\terminate@abbr{\def\last@char{\last}\def\last@space{\ }%
  \@tfor\a@char:=)]'.,?!:;\do{\if\a@char\n@token\def\last@space{}\fi}%
  \if\last.\@tfor\a@char:=.,?!:;\do{\if\a@char\n@token\def\last@char{}\fi}\fi%
  {\sc\last@char}\last@space}
\makeatother
\newcommand{\ket}[1]{\ensuremath{|\,#1\,\rangle}}
\newcommand{\bra}[1]{\ensuremath{\langle\,#1\,|}}
\newcommand{\braket}[2]{\ensuremath{\langle\,#1\,|\,#2\,\rangle}}
\newcommand{\mean}[1]{\ensuremath{\langle{#1}\rangle}}

\newcommand{\card}[1]{\ensuremath{\left|#1\right|}}
\newcommand{\nsygat}{\abbr{n.g.}}
\newcommand{\staerr}{\abbr{s.e.}}
\newcommand{\qc}{\abbr{q.c.}}
\newcommand{\unoq}{\abbr{1q}}
\newcommand{\dueq}{\abbr{2q}}
\newcommand{\id}{\ensuremath{\mathbb{I}}}
\newcommand{\tr}{\ensuremath{\mathrm{tr}}}
\newcommand{\opR}{\ensuremath{\widehat{R}}}
\newcommand{\opS}{\ensuremath{\widehat{S}}}
\newcommand{\opE}{\ensuremath{\mathcal{E}}}
\newcommand{\apc}[1]{^{\,(#1)}}
\newcommand{\expower}[2]{\ensuremath{\mathcal{U}\left(#1,#2\right)}}
\newcommand{\err}{\ast}
\newcommand{\aout}{\ensuremath{\wp}}
\newcommand{\ain}{\ensuremath{\widetilde\wp}}
\newcommand{\myfig}[3][~]{\noindent #1\hfill\begin{minipage}[c]{#2\linewidth}
        \includegraphics[width=.99\linewidth]{#3}\end{minipage}\hfill~\\}
\newcommand{\conline}{(Color online)~}
\makeatletter\def\@fltstk{}\makeatother
\begin{document}
\title{A quantitative model for the effective decoherence \\
        of a quantum computer with imperfect unitary operations}
\author{Stefano Bettelli}
\homepage[Group's home page: ]{http://www.quantware.ups-tlse.fr}
%\email[the author can be reached at:]{bettelli@irsamc.ups-tlse.fr}
\affiliation{Laboratoire de Physique Th\'eorique, FRE 2603 du CNRS,
  Universit\'e Paul Sabatier, 31062 Toulouse Cedex 4, France}
\date{\today}
\preprint{quant-ph/0310152}

\begin{abstract}
  The problem of the quantitative degradation of the performance of a
  quantum computer due to noisy unitary gates (imperfect external control)
  is studied. It is shown that quite general conclusions on the evolution
  of the fidelity can be reached by using the conjecture that the set of
  states visited by a quantum algorithm can be replaced by the uniform
  (Haar) ensemble. These general results are tested numerically against
  quantum computer simulations of two particular periodically driven
  quantum systems.
\end{abstract} 

\maketitle

\section{Introduction}

In the last decade quite a few studies have been carried on which focus on
the behaviour of an imperfect quantum computer (\qc) \cite{ChuangNielsen00}.
Although the sources of imperfections are in general specific to each
particular physical implementation, they can be grouped in two categories
\cite{Preskill98}, {\em decoherence} and {\em unitary errors}.

Decoherence consists in an unwanted interaction be\-tween the \qc and the
surrounding environment, due to imperfect isolation \cite{Zurek03}. This
coupling causes the two parties to become entangled, so that the state of
the \qc alone is no more described by a pure state but by a density matrix.
Its consequences on the results of a computation, which prompted for the
development of error correction strategies, have been studied from the very
early days of quantum computing \cite{early_dec}; active corrections (error
correction codes) are already sufficiently understood to be covered in
standard textbooks \cite[see][chapters 8-10]{ChuangNielsen00}, while
passive methods (decoherence free subspaces) have been developed more
recently \cite[see][for a review]{LidarWhaley03}.

Even in a \qc perfectly isolated from the environment the computation can
still be affected by unitary errors; the state evolution remains coherent,
but the algorithm is slightly modified. Unitary errors arise in at least
two different contexts: they may be due to an imperfect implementation of
the operations by means of which the computation is performed ({\em noisy
  gates}, \nsygat) or to a residual Hamiltonian leading to an additional
spurious evolution of the \qc memory ({\em static errors}, \staerr).

The lack of knowledge about the parameters determining the unitary errors
implies some level of uncertainty for the results of the computation, that
is, unitary errors can be thought of as an {\em effective decoherence}
mechanism. While true decoherence and \nsygat are to a certain extent
similar in their effects, \staerr differ because time correlations cannot
be neglected. This is particularly clear, for instance, in ensemble
computing, where the term {\em incoherence} is used \cite{PBEFFHC03}: even
if the action of \staerr is described by completely positive
superoperators, it can be corrected by locally unitary ``refocussing''
operations.

The effects of imperfections can be summarised by means of a relation
between an ``imperfection intensity'' $\epsilon$ and the timescale $t_d$
for the degradation of some characteristic quantity linked to the
computational process. The usual choice \cite{MPZ97} for this quantity is
the {\em fidelity}, defined as the squared modulus of the overlap between
the state $\ket{\psi} = U\ket{\psi^0}$ of the quantum memory after an ideal
algorithm and the corresponding state $\ket{\psi_\err} = U_\err
\ket{\psi^0}$ after an imperfect evolution of some type,
\begin{equation} \label{eq:fidelity}
  f = \left| \braket{\psi}{\psi_\err} \right|^2.
\end{equation}

There is, recently, a rising interest in the problem of \staerr, especially
in connection to different signatures for integrable/chaotic dynamics
\cite{static_err, BCMS01}. The focus in this article will however be on
\nsygat, which are less explored in the literature. The first
investigations were based on simulations of Shor's algorithm \cite{Shor97}.
In the seminal paper for the ion trap computer \cite{CiracZoller95} the
authors note that the quantum Fourier transform is quite robust with
respect to an imperfect implementation of the laser pulses. A more
systematic but still completely numerical analysis of the evolution of the
fidelity in an almost identical setup was performed in \cite{MPZ97}.
Although only Hadamard gates were considered to be noisy there, the
empirical result captures the general behaviour $t_d \sim \epsilon^{-2}$.

Grover's algorithm \cite{Grover97} was studied in \cite{SongKim00} with the
help of a phenomenological model of probability diffusion, but the
dependence of the model parameters on the number of qubits and $\epsilon$
was not determined. Conclusions similar to those in \cite{MPZ97} were
reached also in \cite{SMB03}, where again the test algorithm was Grover's
one\footnote{Indeed, the result stated by the authors in \cite{SMB03} is
  that there is an error threshold at $\epsilon = O({n_q} ^{-\frac{1}{2}}
  N^{-\frac{1}{4}})$, where $n_q$ is the number of qubits in the \qc and
  $N=2^{n_q}$ is the number of levels.  However, given that Grover's
  algorithm consists of $O(\sqrt{N})$ repetitions of a basic cycle, with
  $O(n_q)$ \nsygat per cycle (only the Hadamard gates are perturbed), it
  turns out that the overall number of \nsygat is $N_g = O(\sqrt{N}n_q)$,
  so that $\epsilon^2 = O(1/N_g)$, i.e.  $t_d \sim \epsilon^{-2}$.}, and in
a number of articles about the simulation of quantum chaotic maps
\cite{BCMS01, SongShepelyansky01, ChepelianskiiShepelyansky02,
  TerraneoShepelyansky03, LGS03}. These results show that the timescale for
the degradation of the fidelity of a \qc decreases only polynomially with
noise and system size\footnote{Although there are other quantities which
  are exponentially sensitive to the number of qubits
  \cite{SongShepelyansky01, LGS03, BettelliShepelyansky03}.}.

The aim of this article is to perform a quantitative analysis of the
effects of {\em unbiased} \nsygat in a \qc running a quantum algorithm 
{\em without measurements} (i.e. a multi-qubit unitary transformation). If
$G$ is an elementary gate accessible to the \qc, its noisy implementation
can be written as $G_{\err} = \opE G$, where $\opE$ is a unitary error
operator (in the following the \mbox{symbol $\err$} refers to the presence
of imperfections); so, an ideal quantum algorithm is $U = \prod_k G_k$ and
a noisy one is $U_\err = \prod_k \opE_k G_k$. The adjective ``noisy''
implies that there is no correlation for the error intensities of $\opE_k$
and $\opE_{k' \ne k}$, and the adjective ``unbiased'' that for an error
operator $\opE = \id + O(\epsilon)$ the average\footnote{Satisfying this
  condition corresponds to tuning the gate implementation in order to
  eliminate the systematic part of the error.} over different realisations
gives $\mean{\opE} = \id + O(\epsilon^2)$.

It will be shown that, in this ``unitary model'', the degradation of the
fidelity for \nsygat depends only on the spectrum of the errors, on the
size of the \qc and on the number of gates in a given algorithm. Note that
in an algorithm not involving measurements, the mean value of the fidelity
is a directly measurable quantity; in fact, due to the previous
considerations, it is sufficient to run the algorithm forward for $N_g/2$
gates, and then to apply the adjoint algorithm to go back to the initial
state (which is known and which can be always chosen as an element of the
computational basis). The average fidelity for $N_g$ gates is then exactly
the probability of measuring this initial state, which can be done at each
fixed precision with a fixed number of measurements.

The theoretical results are checked against numerical simulations based on
two specific quantum algo\-rithms calculating the evolution of a periodically
driven quantum system (see appendix \ref{app:algorithms}), the sawtooth map
\cite{BCMS01} and the double-well map \cite{ChepelianskiiShepelyansky02}.
The former is a purely unitary transformation, while the latter uses an
ancilla and intermediate measurements. These numerical examples concern
thus an imperfect \qc used to simulate the evolution of another quantum
system (with the \qc, in turn, simulated by a classical computer). Note
that the choice of a particular set of primitive gates might affect the
numerical checks but not the theoretical results, since the set is left
unspecified there.

The proposed model and the numerical results suggest that a \qc subject to
a non-trivial computational task shows a sort of ``universal''
(algorithm-independent) behaviour for the fidelity degradation. The body of
the article is organised as follows. Section \ref{sec:uerrors} recalls the
bounds to the fidelity degradation posed by the unitarity of
errors\footnote{ This bounds are valid when the \qc evolution is completely
  unitary, i.e. when the executed algorithm does not use measurements
  before the end of the computation.}. Section \ref{sec:nsygat} introduces
a detailed model for this degradation, whose predictions are tested
numerically in section \ref{sec:numerics}. The final part contains a
discussion of open questions and limitations of this model. Various
appendices explain the details of the calculations, of the implementation
of quantum algorithms and of their numerical simulations.

\section{Fidelity degradation induced by unitary errors: a simple bound}
\label{sec:uerrors}

As already said, a common approach in the study of unitary imperfections is
to analyse their effects by means of the fidelity (equation
\ref{eq:fidelity}), which can be seen as the probability of remaining on
the state selected by the ideal algorithm. In a unitary error model a
simple bound holds for its degradation; in fact, by introducing the squared
norm of the difference of the state vectors one finds
\begin{displaymath}
  \left\| \ket{\psi} - \ket{\psi_\err} \right\|^2
  = 2(1 - \Re\braket{\psi}{\psi_\err}) \geq 2(1-\sqrt{f}).
\end{displaymath}

However, since $f$ does not depend on a global phase change on
$\ket{\psi_\err}$ (i.e. $U_\err\rightarrow e^{i\chi}U_\err$), while
$\left\|\ket{\psi_\err} - \ket{\psi}\right\|$ does, one can take the
minimum over $\chi$. In this case, the previous bound can be shown to
become an equality:
\begin{equation} \label{eq:fidelity2}
  \min_\chi\left\|\ket{\psi}-e^{i\chi}\ket{\psi_\err}\right\|^2
  = 2(1-\sqrt{f}).
\end{equation}

The squared norm can in turn be bounded by the norm of the operator
implementing the algorithm. If the operator is diagonalisable this norm
is of course the largest eigenvalue (in modulus). Introducing
\begin{equation} \label{eq:normOp}
  \| A \| = \max_{\ket{\phi}} \| A\ket{\phi} \|,
\end{equation}
which is normalised to $1$ for unitary operators, it is immediate to see
from definition \ref{eq:normOp} that
\begin{equation} \label{eq:psiplus}
  \left\| \ket{\psi} - e^{i\chi}\ket{\psi_\err} \right\|
  \leq \left\| U - e^{i\chi}U_\err \right\|.
\end{equation}

The inequality $\|U_1U_2 - W_1W_2\| \leq \|U_1 - W_1\| + \|U_2 - W_2\|$,
which holds for unitary operators, implies that the majorisation chain can
be continued with
\begin{eqnarray} \label{eq:Uplus}
  \min_\chi \left\| U - e^{i\chi}U_\err \right\| &\leq& \sum_k
  \min_{\chi_k}\left\| G_k - e^{i\chi_k} \opE_k G_k \right\| \\
  &=& \sum_k \min_{\chi_k} \left\| \id - e^{i\chi_k} \opE_k \right\|
        \stackrel{def}{=} \sum_k \varsigma_{\err k}, \nonumber
\end{eqnarray}
where $\chi$ and the $\chi_k$'s are independent variables, since $\min_\chi
\| \id - e^{i(\chi - \sum_k \chi_k)} \| = 0$. In this expression, the
definition of $\varsigma_{\err k} = \min_{\bar\chi} \left\| \id -
  e^{i\bar\chi} \opE_k \right\|$ has been introduced. This quantity depends
only on the spectrum of the error operator $\opE_k$. This is easily shown
by moving back to definition \ref{eq:normOp} and writing the generic vector
\ket{\phi} over an eigenbasis for $\opE_k$; if the corresponding
eigenvalues are $e^{i\lambda_{kj}}$ ($\lambda_{kj} \in \mathbb{R}$ since
the errors are unitary), one obtains
\begin{eqnarray*}
  \varsigma_{\err k}
  &=& \min_\chi \max_{\ket{\phi}=\sum_j a_{kj}\ket{\phi_{kj}}}
        \left\| (\id - e^{i\chi}\opE_k) \ket{\phi} \right\| \\
  &=& \min_\chi \max_{a_{kj}} \left\|{\textstyle\sum_j}
         a_{kj}(1-e^{i(\chi+\lambda_{kj})}) \ket{\phi_{kj}}\right\| \\
  &=& \min_\chi \max_j \left|1-e^{i(\chi+\lambda_{kj})}\right|
   = 2\min_\chi\max_j \left| \;\sin \textstyle
     \frac{\chi+\lambda_{kj}}{2}\right|.
\end{eqnarray*}

This already shows that only the eigenvalues $e^{i\lambda_{kj}}$ of
$\opE_k$ are relevant. Introducing finally the definition
\begin{equation} \label{eq:defvarsigmaerr}
  \varsigma_{\err} = \left\langle \varsigma_{\err k} \right\rangle,
\end{equation}
where the average for $\varsigma_{\err}$ is on the type and the intensity
of the errors, it is possible to conclude that
\begin{equation} \label{eq:boundU}
  \min_\chi \left\| U - e^{i\chi}U_\err \right\|^2
  \leq \Big( \sum_k \varsigma_{\err k} \Big)^2
  = \varsigma_\err^2 N_g^2.
\end{equation}

Therefore, using equation \ref{eq:fidelity2}, \ref{eq:psiplus} and
\ref{eq:boundU}, and multiplying by $1+\sqrt{f}\leq 2$, one obtains
\begin{equation} \label{eq:general_limit_fidelity}
  1-f \leq \varsigma^2_\err N^2_g,
\end{equation}
which shows that the degradation of the fidelity in a unitary error model
cannot increase more than quadratically in the number of gates and in the
error intensity. Of course, this result is only a majorisation, and the
true average behaviour can be very different.

The quantity $\varsigma_\err$ is in general algorithm dependent, because of
the average over the type and the intensity of the errors, but only through
the spectra of the error operators. In order to calculate $\varsigma_{\err
  k}$, a non-trivial combined minimisation and maximisation is required.
However, if $\opE_k$ is assumed to be a {\em small} error, i.e. if all the
angles $\lambda_{kj}$ are located in a narrow neighbourhood of radius
$r_k$, it can be seen by geometric means that
\begin{equation} \label{eq:intro_rk}
  \varsigma_{\err k} = 2 \sin \frac{r_k}{2} \lesssim r_k.
\end{equation}

\section{Fidelity degradation due to noisy gates: a perturbative model}
\label{sec:nsygat}

The numerical simulations of the earlier articles about unitary errors,
cited in the introduction, show that the degradation of the fidelity
(definition \ref{eq:fidelity}) follows a law $f = \exp(-\alpha N_g
\epsilon^2)$, where $\epsilon$ expresses the ``intensity'' of the
imperfections (see appendix \ref{app:error_model_nsygat} for the details
about the error model underlying \nsygat) and $\alpha$ is a parameter to be
extracted with a fit. Since in this article the focus is on the small error
limit, the goal is to find the value of $\alpha$ for $f \sim 1 - \alpha N_g
\epsilon^2$ (there are however arguments which support the extension to an
exponential law). The generic limit in equation
\ref{eq:general_limit_fidelity} can be specialised for the case of \nsygat
with the value $\varsigma_\err = \epsilon/8$ (see formula
\ref{eq:varsigma_nsygat}):
\begin{displaymath}
  1-f \leq \frac{N^2_g \epsilon^2}{64}.
\end{displaymath}

As already said, the previous relation is only a majorisation. In
particular, equation \ref{eq:Uplus} takes into account the worst case
scenario, where all the \nsygat sum up coherently. One could conjecture
that in a non-coherent scenario formula \ref{eq:boundU} should be modified
by replacing $(\sum_k \varsigma_{\err k})^2$ with $(\sum_k \varsigma_{\err
k}^2)$, which reduces $N_g^2$ to $N_g$ (formula \ref{eq:varsigma_incoherent})
\begin{equation} \label{eq:fidelity4}
  1-f \simeq \frac{N_g \epsilon^2}{48}.
\end{equation}

As already said, exception made for the numerical coefficient, which is
linked to the chosen error model, this result is completely general and can
be found more simply by noting that each noisy gate transfers a probability
of order $\epsilon^2$ to the space orthogonal to \ket{\psi}, and that, in
absence of correlations, all these probabilities can be summed up. The law
in equation \ref{eq:fidelity4} for a noisy computation was first
conjectured and shown numerically in \cite{MPZ97}, where the authors remark
the fact that, although efficient non-trivial circuits in general produce
entanglement in the \qc memory, it seems possible to estimate the
dependence of the fidelity on $\epsilon$ with a model where the \nsygat
errors affect each qubit independently.

This heuristic derivation however leaves two questions unanswered: how to
estimate the exact numerical coefficient and the fluctuations of the
fidelity. In order to answer them, another approach will now be introduced,
dealing with $f$ in the limit of ``small errors''.  In this approach,
$f_k$, the fidelity after $k$ \nsygat's, is treated like a stochastic
variable with $k$-dependent distribution, with the constraint that $f_0$ is
$1$ with certainty.

In the following it will be shown that the effects of errors can be
summarised by a single quantity, the parameter $\sigma^2_\err$, which is
(see also definition \ref{eq:defsigmalambda} for $\sigma_\lambda$) the
average variance (over the error types and intensities) of the phases of
the eigenvalues of the unitary error operators (the limit of small errors
thus corresponds to $\sigma^2_\err \ll 1$):
\begin{equation} \label{eq:sigmastar}
  \sigma_\err^2 = \mean{\sigma^2_{\lambda k}}.
\end{equation}

If \ket{\psi^k} is the state of the quantum memory after $k$ ideal gates,
\ket{\psi_\err^k} that after $k$ noisy gates, and $\mathbb{P}_k =
\ket{\psi^k} \bra{\psi^k}$ the projector onto the ideal subspace, starting
with $\ket{\psi_\err^k} = \opE_k G_k \ket{\psi_\err^{k-1}}$ one easily
shows that
\begin{eqnarray}
  \opE_k^\dagger \ket{\psi_\err^k}
  &=& G_k [\mathbb{P}_{k-1} + (\id - \mathbb{P}_{k-1})]
        \ket{\psi_\err^{k-1}} \nonumber \\
  &=& \ket{\psi^k} \braket{\psi^{k-1}}{\psi_\err^{k-1}} +
        \ket{\phi_\perp^k} \label{eq:fancyevol} \\
  &=& e^{i\theta_{k-1}} \sqrt{f_{k-1}} \ket{\psi^k} +
        \sqrt{1 - f_{k-1}} \ket{\psi_\perp^k}, \nonumber
\end{eqnarray}
where $e^{i\theta_k}$ is a phase such that $\braket{\psi^k}{\psi_\err^k} =
e^{i\theta_k} \sqrt{f_k}$, and $\ket{\psi_\perp^k}$ is the vector
$\ket{\phi_\perp ^k} = G_k (\id - \mathbb{P}_{k-1}) \ket{\psi_\err^{k-1}}$
divided by its norm $\sqrt{1 - f_{k-1}}$; note that \ket{\psi^k_\perp}
lives in the space orthogonal to \ket{\psi^k }. For shortness of notation let
\begin{eqnarray*} 
  \aout_k &=& e^{-i\zeta_k} \bra{\psi^k}\opE_k\ket{\psi^k}
        \label{eq:defaoutain} \\ \mathrm{and}\quad 
  \ain_k &=& e^{-i\zeta_k}e^{-i\theta_{k-1}} \bra{\psi^k}
        \opE_k\ket{\psi^k_\perp},
\end{eqnarray*}
with $e^{i\zeta_k} = \lim_{\,\sigma_\err \rightarrow 0} \bra{\psi^k} \opE_k
\ket{\psi^k}$. The introduction of $\zeta_k$ is motivated by the fact that
in this way $\aout_k$ goes to $1$ in the limit of small errors, instead of
depending on a global phase for $\opE_k$. In other words, $\ln\opE_k =
i(\zeta_k\id + Q_k)$, where $Q_k = O(\sigma_\err)$ is an Hermitian matrix.
Multiplying on the left both sides of equation \ref{eq:fancyevol} by
$\bra{\psi^k} \opE_k$ and taking the squared modulus one finds
\begin{equation} \label{eq:master_fk}
  f_k = \left| \sqrt{f_{k-1}} \cdot \aout_k +
  \sqrt{1 - f_{k-1}} \cdot \ain_k \right|^2.
\end{equation}
$f_k$ is therefore determined by two competing contributions: the first
term is a loss of fidelity due to the noisy gates moving probability out of
the subspace of \ket{\psi^k}; the second term represents interference from
the subspace where \ket{\psi^k_\perp} lives. 

The quadratic forms $\aout$ and $\ain$, which, in absence of errors, are
equal respectively to $1$ and $0$, have a magnitude which depends on
$\sigma_\err$. By replacing $\ln\opE_k = i(\zeta_k\id + Q_k)$ it is easy to
see that both $1-|\aout|^2$ and $|\ain|^2$ must be $O(\sigma_\err^2)$.
Expanding the squared modulus in equation \ref{eq:master_fk} one finds the
following recursive relation, where the under-scripts stand for the orders
of magnitude of the leading terms with respect to $\sigma_\err$:
\begin{eqnarray} 
  f_k - f_{k-1} &=& \underbrace{f_{k-1}}_{0,2} \underbrace{(|\aout_k|^2-1)}_2
        + \underbrace{(1-f_{k-1})}_2 \underbrace{|\ain_k|^2}_2 \nonumber \\
  && + 2 \underbrace{\sqrt{f_{k-1}}}_{0,2} \underbrace{\sqrt{1-f_{k-1}}}_1
        \Re(\underbrace{\aout_k^{\phantom{*}}}_{0,1} \underbrace{\ain_k^*}_1).
  \label{eq:fidnohyp} \qquad
\end{eqnarray}

Now, keeping only the second order terms in formula \ref{eq:fidnohyp},
taking the limit of the zero order terms for $\sigma_\err \rightarrow 0$,
and summing up the partial differences, one arrives at
\begin{equation} \label{eq:fidhyp1}
  f_{N_g} = 1 + \sum_{k=1}^{N_g} \left[ (|\aout_k|^2-1)
        + 2 \sqrt{1-f_{k-1}} \, \Re\ain_k \right] + O(\sigma_\err^3).
\end{equation}

The first interesting quantity to be calculated from equation
\ref{eq:fidhyp1} is the fidelity averaged over many realisations of noise,
$\mean{f_{N_g}}$. Since the error intensities for different $\opE_k$'s are
uncorrelated, the overall average splits into averages at fixed $k$. The
$\ain_k$ term feels this average almost only because of $\opE_k$, because
the vector \ket{\psi^k_\perp} is determined mainly by the previous history
of the evolution (i.e. the previous, uncorrelated errors). Since errors are
unbiased, one gets $\mean{e^{-i\zeta_k} \opE_k} = \id + O(\sigma_\err^2)$,
so that $\mean{\ain_k} = O(\sigma_\err^2)$ can be neglected\footnote{It is
  not evident that this cancellation holds for \staerr or for biased
  \nsygat; this is the reason why $1 - \mean{f_{N_g}}$ can be proportional
  to $N_g^2$ instead of $N_g$. The similarities between biased \nsygat and
  \staerr have not been inspected so far.} in equation \ref{eq:fidhyp1}.
One is therefore left with
\begin{equation} \label{eq:fidhyp2}
  \mean{f_{N_g}} =
  1 + \sum_{k=1}^{N_g} \mean{ |\aout_k|^2-1 } + O(\sigma_\err^3).
\end{equation}

This leading-order model returns a fidelity which is a function of the
algorithm, because $\aout_k$ depends on \ket{\psi^k}. In a different
approach, one could consider the algorithm itself as another random
variable, so that $\aout_k$ and $\ain_k$ would depend on three sources of
``randomness'':
\begin{description}
\item[\it\underline{the error type}:] the parametric form of $\opE_k$,
  linked to the gate type $G_k$; this allows different elementary gates to
  be affected by different types of errors;
\item[\it\underline{the error intensity}:] the parameter which controls the
  magnitude of the error in $\opE_k$; this allows errors for different
  realisations of the same elementary gate to be different; its
  distribution is crucial for an error model being unbiased or not;
\item[\it\underline{the state vectors}:] among which the \ket{\psi^k}'s
  depend on the history of the ideal algorithm and the \ket{\psi_\perp^k}'s
  on its noisy realisation.
\end{description}

In appendix \ref{app:averages} it is shown how to calculate the average
value of $|\aout|^2$ when \ket{\psi^k} is replaced by a vector of norm $1$
randomly chosen according to the Haar (uniform) measure (this distribution
can be generated by applying the unitary circular ensemble \cite{Mehta91}
to a fixed vector). It turns out that the distribution of $|\aout|^2$ is
exponentially peaked around its average value; the variance is of order
$O(1/\sqrt{N})$, where $N=2^{n_q}$ is the number of levels in the \qc
memory and $n_q$ is the number of qubits. This means that, for large $N$,
if $|\aout|^2$ is calculated on a random vector, the result is a value
exponentially close to the mean with high probability: this can be
rephrased by saying that ``almost all vectors are typical''.

This result is derived using the uniform distribution, but it can be argued
that in order to radically change this picture, the distribution of the
$\ket{\psi}$'s should be very different from the uniform one. This indeed
can happen if the state of the system at some point factorises on the
single qubits; given that the elementary gates are likely to be local to
qubits, the effective $N$ in this case would be very small (for instance,
it would be $2$ for \unoq gates), invalidating the {\em concentration of
  measure} \cite{conc_measure} result. However, it is common wisdom that
efficient algorithms generate state sequences with non-negligible
multi-qubit entanglement. So, one can be led to think that ``interesting''
algorithms satisfy very well the ``typical vector'' assumption. This fact
could have some relation with the recent result that a circuit with a
polynomial number of gates is able to approximate the statistical
properties of the uniform ensemble, although the latter is described by an
exponential number of parameters \cite{Emerson03}.

For these reasons, it will be conjectured here that the result of equation
\ref{eq:fidhyp2} can be approximated by having each $|\aout_k|^2$ replaced
by the average value of $|\aout|^2$ according to the uniform measure. This
replacement, of course, flushes away any dependence on the algorithm due to
the actual sequence of the \ket{\psi^k} vectors; it will be indicated in
the following by an underline. So, the symbol $\mean{~}$, like in
$\sigma_\err^2 = \mean{\sigma^2_{\lambda k}}$, implies only an average over
the error type ($\rightarrow k$) and intensity. Of course, the ``uniform
average'' must be taken on the actual function of $\aout$ or $\ain$ instead
of calculating the function on the average of these variates; in other
words, it is consistent, for instance, to replace $\ain$ with zero but
retain $|\ain|$.

Due to the aforementioned assumption $|\aout_k|^2$ can be replaced by
$\underline{ |\aout_k|^2} = 1 - A \sigma^2 _{\lambda k} +O(\sigma_{\lambda
  k}^4)$ (see formula \ref{eq:average_small}, where $A = \,\scriptstyle
N/(1+N)$ is a vanishing dependence on the dimension of the state space very
close to 1); formula \ref{eq:fidhyp2} becomes then
\begin{equation} \label{eq:fidelity-mean}
  1 - \mean{f_{N_g}} + O(\sigma_\err^3)
  = A \sum_{k=1}^{N_g} \mean{ \sigma^2_{\lambda k} } = AN_g\sigma_\err^2;
\end{equation}
this result agrees with formula \ref{eq:fidelity4} but for the numerical
factor (which has however the correct order of magnitude since
$\scriptstyle \frac{\epsilon^2}{64} \leq \sigma_\err^2 \leq \frac{
  \epsilon^2}{48}$, according to formula \ref{eq:sigma2err-specific}) and
the slight dependence on $N$ given by $A$ (hard to prove numerically).

The second interesting quantity to be calculated is the variance $\sigma^2
(f_{N_g})$. It is difficult to extract the exact variance from the model in
equation \ref{eq:fidhyp1}; however, since it is used only to decide the
number of retries for numerical simulations, an approximate solution is
acceptable. If $\sigma(f_{N_g}) \ll 1 - \mean{f_{N_g}}$ then $f_{k-1}$ in
formula \ref{eq:fidhyp1} can be replaced by its average value given by
equation \ref{eq:fidelity-mean}. The simplified model becomes then
\begin{equation} \label{eq:fidhyp3}
  f_{N_g} \sim 1 - \sum_{k=1}^{N_g} \left[ (|\aout_k|^2 - 1) + 2
  \sigma_\err \sqrt{A(k - 1)} \,\Re\ain_k \right].
\end{equation}

Forgetting for a moment the average on the algorithm, the variance of the
fidelity can be calculated using the identity $\sigma^2(y) = \sum_k
\sigma^2(x_k) + \sum_{k\ne h}\mathrm{cov}(x_k,x_h)$, valid for a variate $y
= \sum_k x_k$. When this identity is applied to formula \ref{eq:fidhyp3},
only the correlations between the terms with the same gate index $k$
survive; therefore, setting the factor $A$ to $1$ (which it is
exponentially close to), and confusing $N_g\pm1$ with $N_g$ where not
sensitive, one finds
\begin{eqnarray}
  \sigma^2(f_{N_g})
  &\simeq& {\textstyle\sum_k} \sigma^2(|\aout_k|^2)
    + 4\sigma_\err^2 {\textstyle\sum_k} (k-1)\sigma^2(\Re\ain_k) \nonumber\\
  && + 4\sigma_\err {\textstyle\sum_k} \sqrt{k-1}
    \cdot \mathrm{cov}(|\aout_k|^2, \Re\ain_k ) \nonumber \\
  &\simeq& N_g \mean{\sigma^2(|\aout_k|^2)}
    + 2 N_g^2 \sigma_\err^2 \cdot \mean{\sigma^2(\Re\ain_k)} \nonumber \\
  && + {\scriptstyle\frac{8}{3}} N_g^{3/2} \sigma_\err \cdot
        \mean{\mathrm{cov}(|\aout_k|^2, \Re\ain_k)} \label{eq:sigma2fn} 
\end{eqnarray}

Note that the three terms in the previous expression are proportional
respectively to $N_g^{-1}\varepsilon^2$, $\varepsilon^2$ and $N_g^{-1/2}
\varepsilon^2$, where $\varepsilon = N_g\sigma_\err^2$; therefore, for a
large number of gates it is the second term (that with the variance of
$\Re\ain_k$) which dominates. Since the parameter which expresses the
difficulty of extracting $\sigma_\err$ from a single numerical simulation
is the ratio between the fidelity standard deviation $\sigma(f_{N_g})$ and
the average fidelity decrease $1-\mean{f_{N_g}} \propto \varepsilon$, a
natural approximation is to keep only the term scaling as $N_g^0
\varepsilon^2$ in formula \ref{eq:sigma2fn}. So, for $N_g \gg 1$ formula
\ref{eq:sigma2fn} becomes
\begin{displaymath}
  \sigma^2(f_{N_g}) \simeq 2N_g^2\sigma_\err^2
  \cdot \mean{\sigma^2(\Re\ain_k)} = 2 N_g^2 \sigma_\err^2
  \left(\mean{\Re^2\ain_k} - \mean{\Re\ain_k}^2 \right). \hspace{-1em}
\end{displaymath}

As before, the average over the error realisations for each $\ain_k$ is
null, therefore $\mean{\Re\ain_k} \rightarrow 0$. Then, expanding the real
part the previous formula becomes
\begin{displaymath}
  \sigma^2(f_{N_g}) \simeq \frac{1}{2} N_g^2 \sigma_\err^2
  \mean{\ain_k^{\,2} + \ain_k^{\,*2} + 2|\ain_k|^2 }
\end{displaymath}

The uniform average prescription for $\ain_k$ consists in choosing
\ket{\psi^k} uniformly random and \ket{\psi^k_\perp} uniformly random in
the orthogonal subspace. As shown in appendix \ref{app:averages}, the term
$\ain_k^2$ averages to zero, and the term $|\ain_k|^2$ can be replaced by
$\underline{|\ain_k|^2} = \frac{N}{N^2-1} \sigma_{\lambda k}^2$ (formula
\ref{eq:av_ain_m2}). Approximating $\frac{N}{N^2-1} \mapsto \frac{1}{N}$
just like $A \mapsto 1$, it is possible to conclude that
\begin{eqnarray} 
  \sigma^2(f_{N_g}) \nonumber
  &\simeq& N_g^2\sigma_\err^2\cdot\mean{\underline{|\ain_k|^2}}
  = N_g^2\sigma_\err^2 {\textstyle\frac{1}{N}}\mean{\sigma_{\lambda k}^2} \\
  &=& {\textstyle\frac{1}{N}} (N_g\sigma_\err^2)^2
  = {\textstyle\frac{1}{N}} (1 - \mean{f_{N_g}})^2. \label{eq:fidelity-sigma}
\end{eqnarray}
The ratio between the fidelity standard deviation $\sigma(f_{N_g})$ and the
average fidelity decrease $1 - \mean{f_{N_g}}$ depends therefore only on
$n_q$ (remember that the number of gates is $N_g$ while the size of the \qc
memory is $N=2^{n_q}$).

Two general remarks should be made on this detailed fidelity model. First,
strangely enough, in equation \ref{eq:fidhyp1} it is the first term which
determines the average fidelity, but the second one which governs its
fluctuations. If the contribution of $\ain$ in relation \ref{eq:sigma2fn}
was null, then $\sigma^2(f_{N_g}) \propto N_g$ and the ratio in formula
\ref{eq:fidelity-sigma} would drop to zero as $N_g^{-1}$, i.e. the fidelity
evolution would be self-averaging; moreover, $f_{N_g}$ would decrease
monotonically as a function of $N_g$; both these consequences are
contradicted by numerical simulations (see, for instance, figure
\ref{fig:nsygat-gamma}).

Second, in the derivation of the statistical properties of $f_{N_g}$ the
number of gates $N_g$ was regarded as a constant. This does not mean that
the fidelity values at different stages of the algorithm are not
correlated. Indeed, this correlation can be large, because of the limit
on the fidelity decay discussed in section \ref{sec:uerrors}.

In the specific error model studied in this article (see appendix
\ref{app:error_model_nsygat}), $\sigma^2_\lambda$ depends only on the
number of qubits which the error operators act on (\unoq or \dueq errors).
For $R_\nu(\xi)$ the eigenvalues are $\{e^{-\xi/2},e^{\xi/2}\}$, and one
gets $\sigma^2_\lambda = \xi^2/4 + O(\epsilon^4)$. On the other hand, for
controlled phase shift errors, the eigenvalues are $\{1,1,1,e^{i\xi}\}$ and
the variance is $\sigma^2_\lambda = 3\xi^2/16 + O(\epsilon^4)$. Thus, in
general, $\sigma^2_\lambda = C\xi^2 + O(\epsilon^4)$, with $C$ depending
only on the gate being \unoq or \dueq and $\xi$ only on the error parameter
distribution. The mean value $\mean{\sigma^2_{\lambda k}}$ implies both an
average on the error type and intensity, giving
\begin{equation} \label{eq:sigma2err-specific}
  \sigma_\err^2 = \mean{\sigma^2_{\lambda k}} = \mean{C}\mean{\xi^2}
  = \frac{n_1C_1 + n_2C_2}{12} \,\epsilon^2,
\end{equation}
where $0\leq n_1(n_2) \leq 1$ is the average number of \unoq (\dueq) gates
during the computation (so that $n_1+n_2=1$). Since $C_1 = \frac{1}{4}$ and
$C_2 = \frac{3}{16}$, this implies $48 \leq (\epsilon^2 / \sigma_\err^2)
\leq 64$.

\section{Numerical checks} \label{sec:numerics}

\begin{figure}[t]
  \myfig[(a)]{.8}{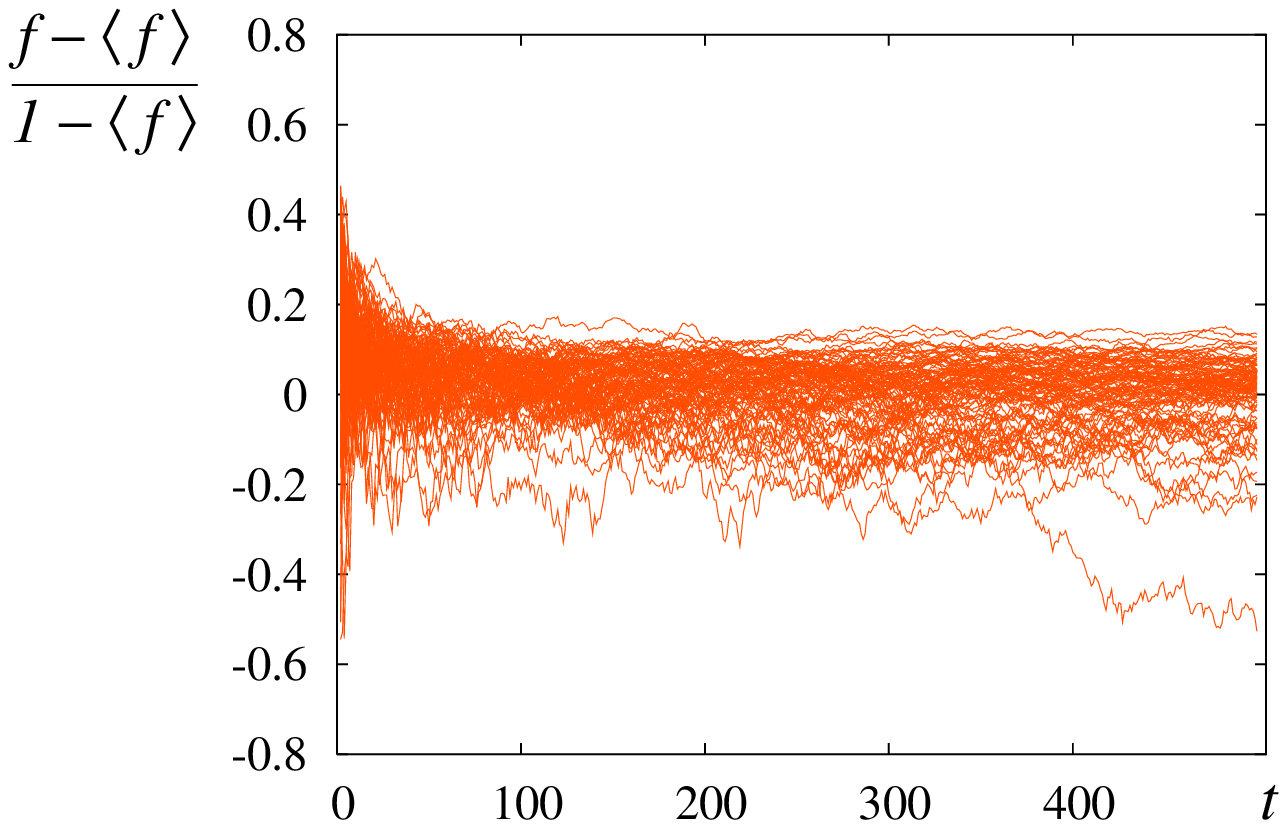}
  \myfig[(b)]{.8}{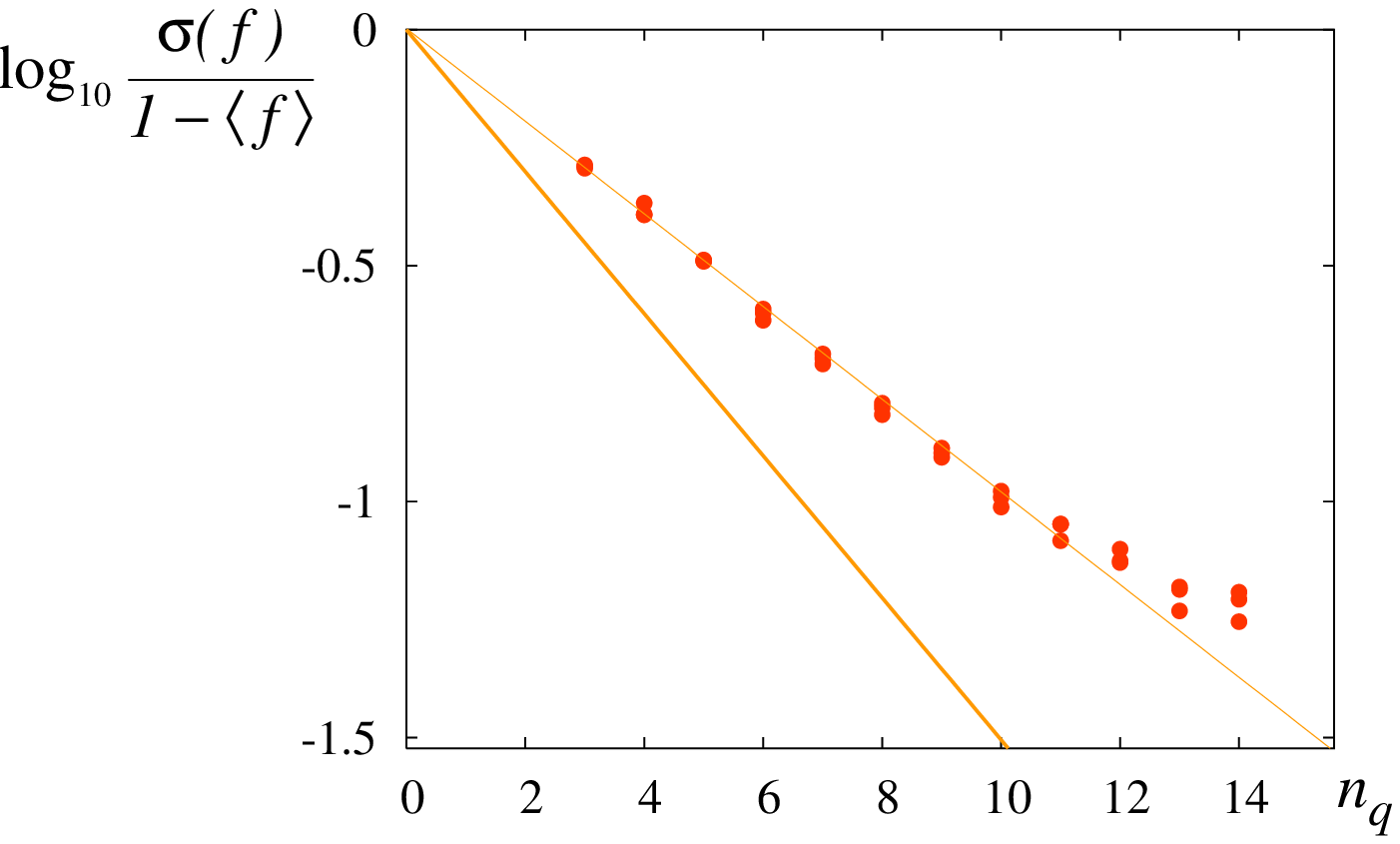}
  \caption[Fidelity fluctuations for the sawtooth map]{
    \label{fig:nsygat-sig-st} \conline \\
    %%As in the following, all plotted quantities are dimensionless. \\
    {\bf(a)} Ratio $\frac{f-\mean{f}}{1-\mean{f}}$, where $f$ is the
    unaveraged fidelity and $\mean{f}$ is its theoretical value given by
    formula \ref{eq:fidelity-mean}, for $100$ numerical simulations on a
    sawtooth map with $n_q=12$ and $\epsilon=0.002$. As in the other
    numerical simulations in this article involving the sawtooth map, the
    number of cells is $L=2$, the classical parameter is $K=0.04$, the
    fidelity is observed in position representation and the initial state
    is $H \otimes \id \otimes H^{\otimes (n_q-2)} \ket{0}$; see appendix
    \ref{app:algorithms} for further details. The same behaviour
    (a spread depending only on the number of qubits for $N_g \gg 1$)
    was found for all the numerical simulations used in figure
    \ref{fig:nsygat-sig-st}-b. The largest fluctuations tend to be located
    below the average. \\
    {\bf(b)} Values of $\frac{\sigma(f)}{1-\mean{f}}$, each point
    corresponding to $676$ numerical simulations at fixed error intensity
    $\epsilon$ and number of qubits $n_q$ ($\mean{f}$ is the numerical
    average of $f$). For each $n_q$ in $\{3,\dots, 14\}$, the error
    intensities $\epsilon = 0.002$, $0.005$ and $0.01$ were tried. Formula
    \ref{eq:fidelity-sigma} predicts that these points depend only on
    $n_q$. The exact dependence $\sqrt{1/N}$, thick line, is however poorly
    followed. The thin line shows a fit with an exponential function
    $aN^{-b}$, giving $a\sim 1$ and $b\sim 1/3$.
  }
\end{figure}

The described fidelity decay model has been tested numerically on two
algorithms simulating periodically driven quantum system: the sawtooth map
\cite{BCMS01} and the double-well map \cite{ChepelianskiiShepelyansky02}.
These systems are usually studied in the quantum chaos field, because in
spite of their relative simplicity they show a large variety of physical
phenomena, from dynamical localisation to quantum ergodicity, first studied
on the kicked rotator model \cite{maps_phenomena}. If $\hat{\eta}$ and
$\hat{\theta}$ are momentum and coordinate operators, then the evolution
corresponding to one map step for a map with potential $V(\theta)$ is given by
\begin{displaymath}
  \hat{U}_F = \exp\left(-i\frac{\hat{\eta}^2}{2\hbar}T\right)
       \exp\left(-i\frac{kV(\hat{\theta})}{\hbar}\right)
\end{displaymath}
(see formula \ref{eq:opFloquet} and other physics and implementation
details in appendix \ref{app:algorithms}). The behaviour of the classical
counterparts of these chaotic maps is controlled by a single parameter
$K=kT$, which in the following simulations is set to $0.04$. For this
value, the double-well map presents a mixed phase space, with a large
island of integrable motion in a chaotic sea; the sawtooth map is chaotic
for every $K>0$, but the cantori regime extends up to $K \sim 1$, so that
the presented numerical data best describe a perturbative regime. In both
cases, it is not evident that the success of the uniform measure
approximation depends on ``chaos''.

The algorithms corresponding to the simulation of these maps contain the
Fourier transform as a basic ingredient for changing the representation
from coordinate to momentum and back. Since the simulated maps have
periodic impulsive kicks, it is possible to write the overall evolution as
a piecewise diagonal unitary transformation in the appropriate
representation. The set of elementary gates which are used for the
algorithms implementation contains single and double qubit phase shifts and
Hadamard gates (see appendix \ref{app:error_model_nsygat} for more
details). Since these algorithms are periodic, the results are usually
expressed in terms of the number of gates $n_g$ per maps step. In the
following the variable $t$ will be used when referring to the ``map time''
(the number of map steps). It is immediate to specialise the previous
results for a map simply by replacing $N_g$ with $n_g t$. The effective
decoherence parameter $\Gamma$ is defined by $f \sim 1 - \Gamma t$.

\begin{figure}[t]
  \myfig{.8}{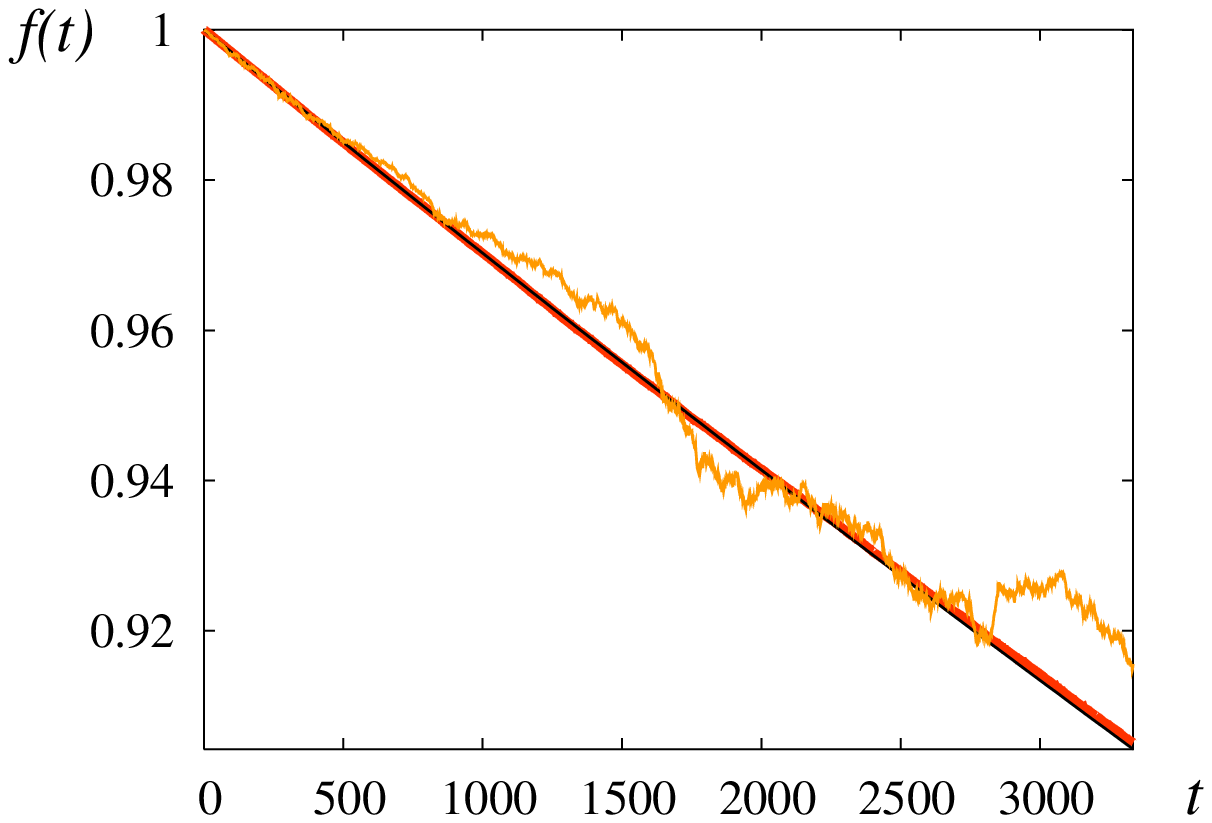}
  \caption[Fidelity degradation in the sawtooth map - example]{
    \label{fig:nsygat-gamma-st-example} \conline
    Decay of the fidelity due to \nsygat during the simulation of the
    sawtooth map, with $n_q=10$ and $\epsilon=0.003$. The simulation of one
    map step for such $n_q$ requires $n_g=200$ gates, with $n_1=45$ \unoq
    gates and $n_2=155$ \dueq gates. The value of $\Gamma$ from formula
    \ref{eq:fidelity-mean} is therefore $\simeq \epsilon^2 n_g / 59.6$,
    shown by the straight line. The red (dark) curve corresponds to an
    average over $500$ numerical simulations (it is hardly distinguishable
    from the straight line in this picture), while the orange (light) curve
    to a single try. See figure \ref{fig:nsygat-sig-st}-a for the map
    parameters.}
\end{figure}

As a first check, the theoretical predictions were compared with the
results of the simulation of a sawtooth map. This map is similar to the
double-well map but it is free from the complications of using auxiliary
qubits. Figure \ref{fig:nsygat-sig-st} is a check of formula
\ref{eq:fidelity-sigma}; it shows that indeed the ratio $\sigma(f_{N_g}) /
(1-\mean{f_{N_g}})$ depends only on $n_q$ and is exponentially small with
it, but the theory predicts a decrease like $N^{-1/2}$ while $N^{-1/3}$ is
observed. A possible explanation could be that the vectors \ket{\psi^k
  _\perp} do not explore the complete $(N-1)$-dimensional space orthogonal
to \ket{\psi^k}, but only $\sim N^{2/3}$ dimensions.

\begin{figure}[p]
  \myfig[(a)]{.8}{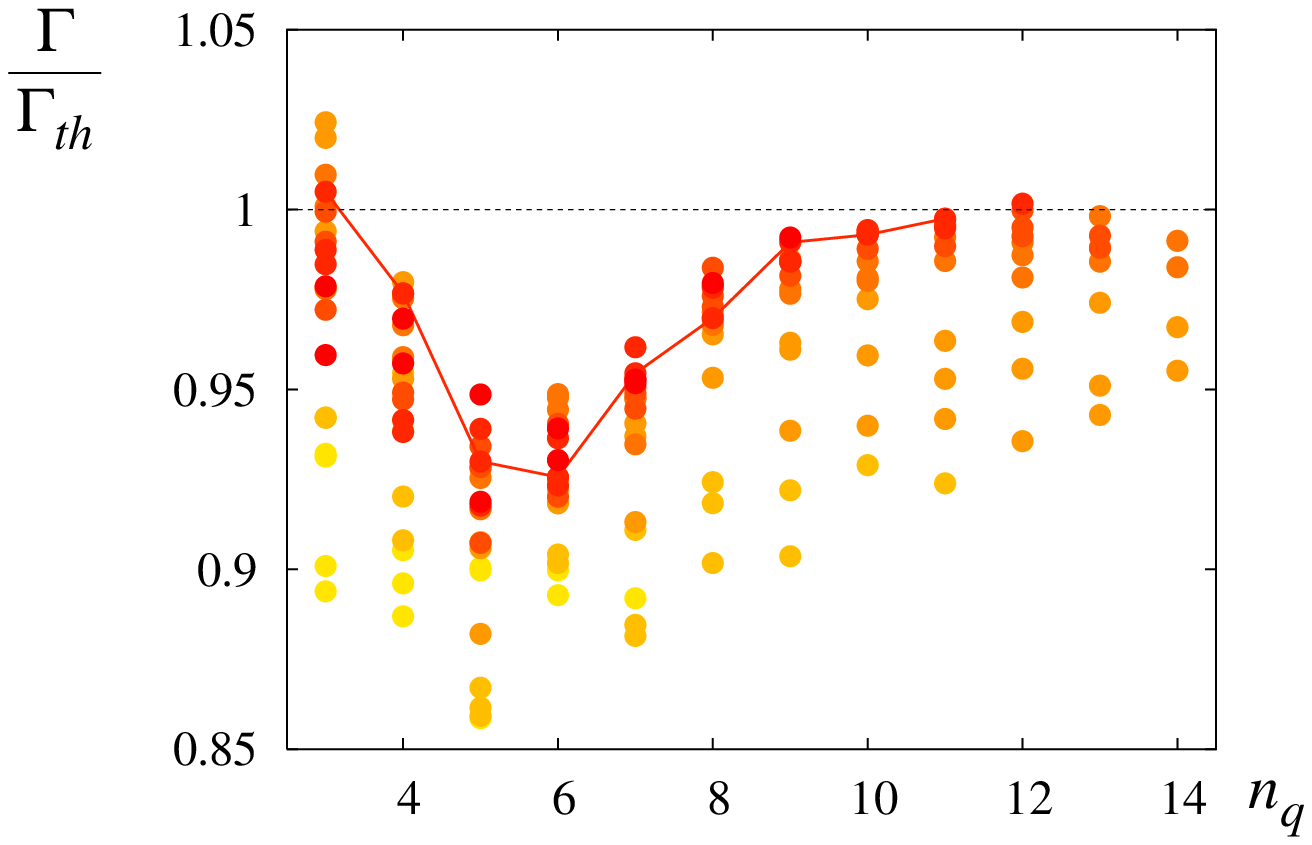}
  \myfig[(b)]{.8}{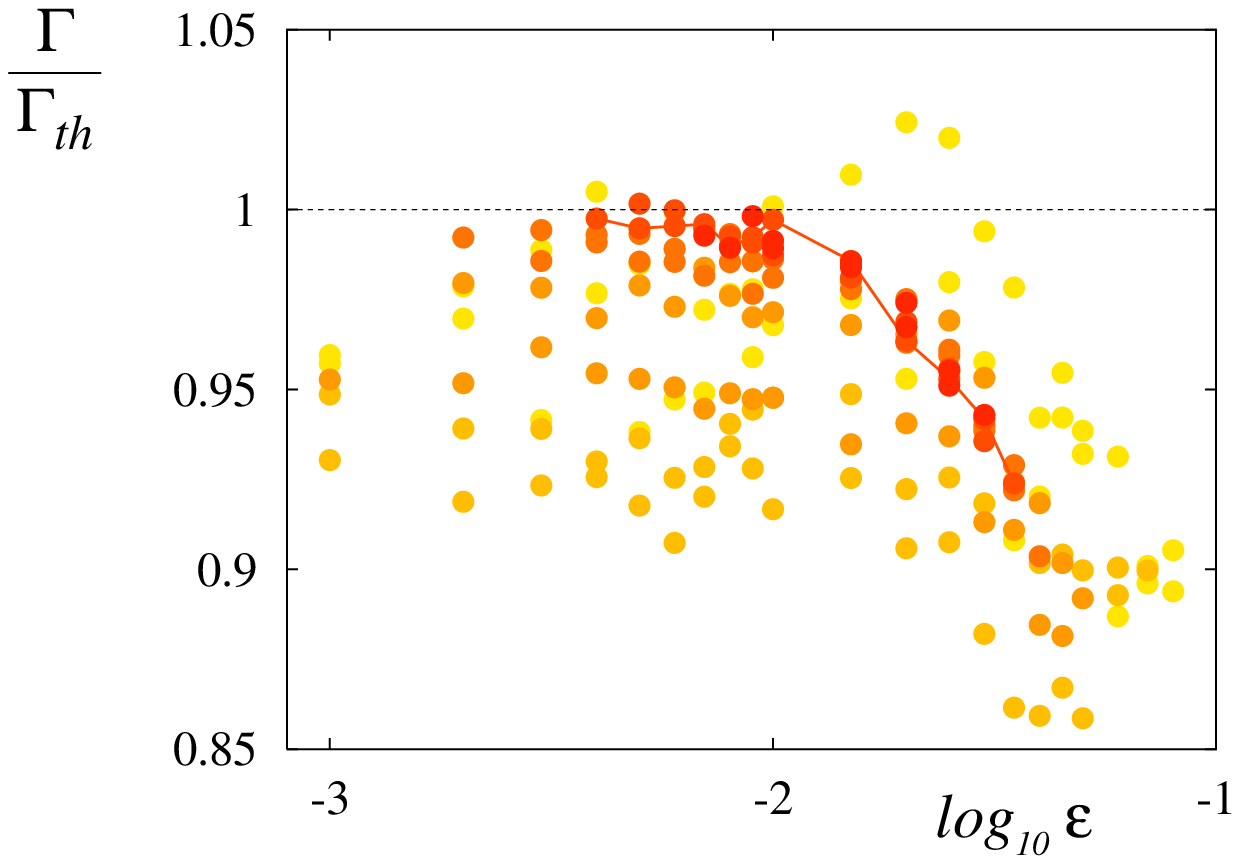}
  \myfig[(c)]{.8}{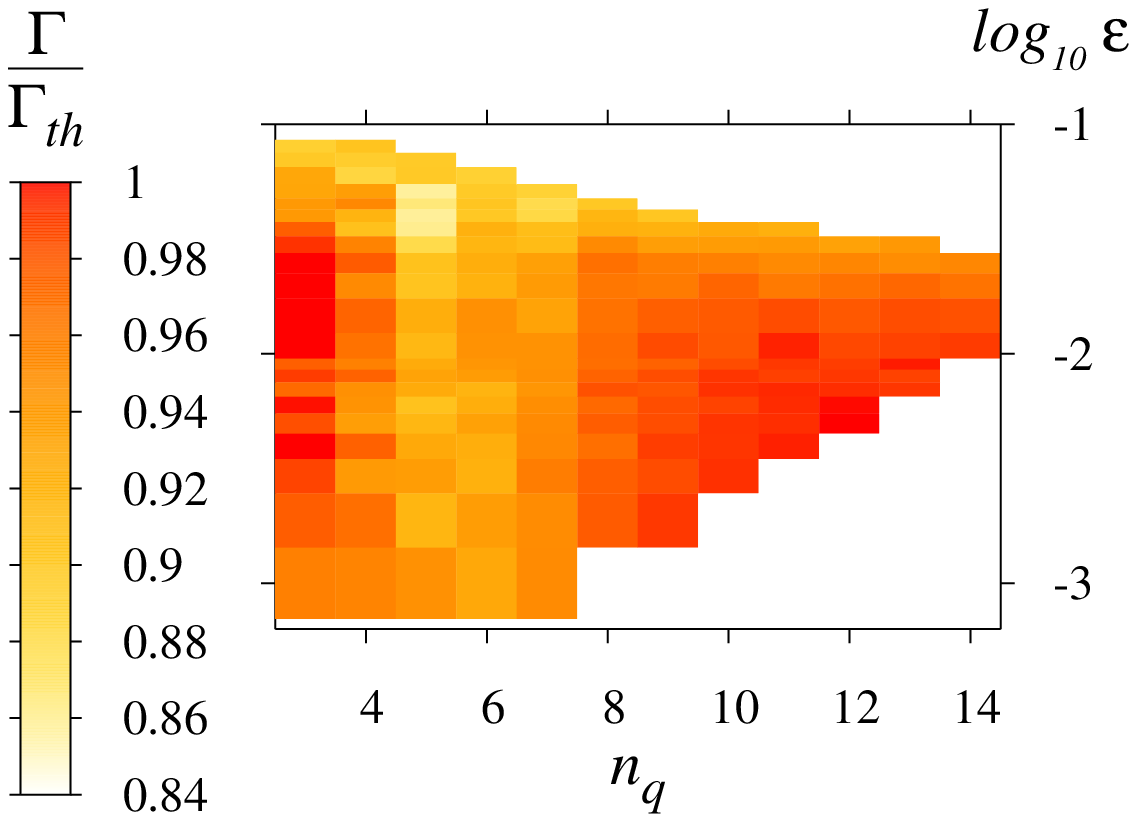}
  \caption[Fidelity degradation in the sawtooth map]{
    \label{fig:nsygat-gamma-st} \conline \\
    {\bf(a)} Ratio $\Gamma/\Gamma_{th}$ versus number of qubits $n_q$ for
    the sawtooth map. $\Gamma$ is the (numerical) fidelity decay constant
    and $\Gamma_{th}$ is its theoretical average, given by formula
    \ref{eq:fidelity-mean}. Each point corresponds to an average at fixed
    $n_q$ and $\epsilon$ with $500$ numerical simulations.  Yellow (light)
    points correspond to large errors ($\epsilon \sim 10^{-1}$) while red
    (dark) points to small errors ($\epsilon \sim 10^{-3}$). The solid line
    is an eye-guide for $\epsilon = 0.004$. See figure
    \ref{fig:nsygat-sig-st}-a for the map parameters. \\
    {\bf(b)} Same as picture \ref{fig:nsygat-gamma-st}-a, but the
    independent variable is the error intensity $\epsilon$, while the
    yellow to red (light to dark) transition corresponds to an increasing
    number of qubits. The solid line is an eye-guide for $n_q = 11$. See
    figure \ref{fig:nsygat-sig-st}-a for the map parameters. \\
    {\bf(c)} Same as picture \ref{fig:nsygat-gamma-st}-a/b, but the
    dependence on the error intensity $\epsilon$ and the number of qubits
    $n_q$ is shown at the same time. Each coloured box corresponds to an
    average at fixed $n_q$ and $\epsilon$ with $500$ numerical simulations.
    The yellow to red (light to dark) transition corresponds to an
    increasing value of the ratio, which remains however in general limited
    by $1$. Missing boxes are due either to combinations taking too long to
    simulate or to fidelities dropping too fast. See figure
    \ref{fig:nsygat-sig-st}-a for the map parameters.
  }
\end{figure}

The previous ratio characterises the difficulty of fitting $\Gamma$ from a
single numerical simulation, and can be used to estimate the number
$n_{sim}$ of tries necessary to reduce the statistical uncertainty, for
instance setting $\frac{1}{\sqrt{n_{sim}}}\frac{\sigma(f)}{1-\mean{f}} \sim
1\%$. This prescription implies a number of simulations $n_{sim} \sim 10^4
2^{-n_q}$; however, it is better to use the numerical values from figure
\ref{fig:nsygat-sig-st}-b, because the theoretical formula gives an
underestimate. In the remaining of the article $n_{sim}$ varies between
$50$ to $500$.

An example of the degradation of the fidelity for the sawtooth map is given
in figure \ref{fig:nsygat-gamma-st-example}, where it can be clearly seen
that a single numerical simulation can give a fidelity which is not
monotonic and correlated over many map steps. The prediction $\Gamma = A
n_g \sigma_\err^2$, given by formula \ref{eq:fidelity-mean}, is compared
with numerical simulations, again for the sawtooth map, in figure
\ref{fig:nsygat-gamma-st}. It turns out that the theoretical value is an
upper bound for the the numerical fidelity decay constant, and that, not
surprisingly, the agreement is better for a large number of qubits and
small errors; for $n_q \gtrsim 10^1$ and $\epsilon < 10^{-2}$, the
discrepancy is within $5\%$.

\begin{figure}[t]
%  \myfig{.8}{FIG-nsygat-averages.eps}
  \myfig{.8}{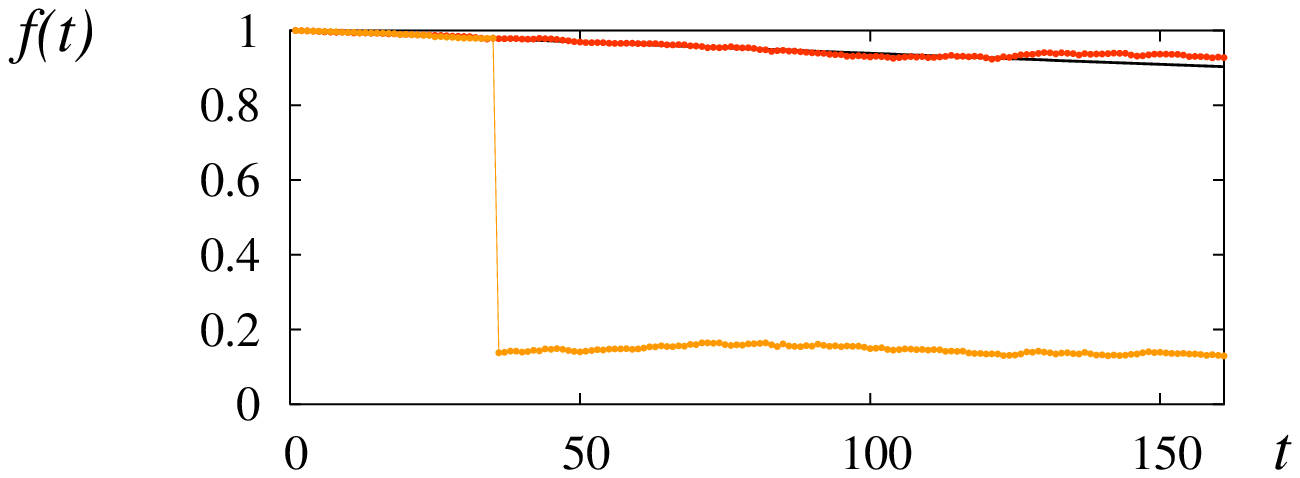}
  \caption[Measurements in the double-well map simulation]{
    \label{fig:nsygat-anomalous} \conline %\\
%    Decay of the fidelity due to \nsygat during the simulation of the
%    double-well map, with $n_q=5$ and $\epsilon=0.01$. Since the algorithm
%    uses an ancilla, the number of levels is $N=2^{n_q+1}=64$. The
%    simulation of one step of the map requires $n_g=362$ gates, with $153$
%    \unoq gates and $209$ \dueq gates.  The value of $\Gamma$ from formula
%    \ref{eq:fidelity-mean} is therefore $\simeq \epsilon^2n_g/57$, shown by
%    the straight line. The red (dark) curve corresponds to an average over
%    $50$ numerical simulations, while the orange (light) curve to a single
%    numerical simulation (note that it is not monotonic and correlated over
%    many steps). \\
    Comparison of a typical and anomalous decay of the fidelity
    for the double-well map, with $n_q=5$ (plus one ancilla) and $\epsilon
    = 0.01$. Both curves correspond to a single numerical simulation. The
    anomalous drop is due to a measurement of the ancilla giving the
    ``wrong'' result (note the change in the vertical scale). As in the
    other numerical simulations, unless otherwise stated, the number of
    cells is $L=2$, the classical parameter is $K=0.04$, the centre of the
    potential well is $a=1.6$, the fidelity is observed in position
    representation and the initial state is $H\otimes\id\otimes H^{\otimes
    (n_q-2)} \ket{0}$ (see appendix \ref{app:algorithms}).
  }
\end{figure}

The simulation of the fidelity decay on the double well map presents a new
effect: beyond ``standard'' fidelity fluctuations, there are some rare
``anomalous'' fluctuations, like that in figure \ref{fig:nsygat-anomalous},
with the fidelity dropping abruptly to almost zero. This is due to the fact
that the algorithm implementation, for $n_q \ge 4$, uses an auxiliary qubit
(an ancilla), which is reinitialised to \ket{0} every time it is reused;
the reinitialisation implies a measurement, which, when the evolution is
affected by noise, can select the ``wrong'' result (i.e. \ket{1}), thus in
practice completely destroying the computation. These catastrophic events
become more and more frequent while $\epsilon$ increases.

However, when the measurement of the ancilla gives the ``good'' result, it
turns out that the degradation of the fidelity is slowed down. This
cancellation of the (unwanted) evolution is known as Zeno or watchdog
effect \cite{zeno_origin}; recently it has been shown that this inhibition
of decoherence can arise in more general contexts, where the essential
ingredient is a strong coupling to the environment and not its trivial
dynamics \cite{zeno_general}. In the simulation of the double-well map it
is indeed the ancilla which plays the role of the environment.

The repeated observation of the ancilla can be thought of as an error
correction strategy in the Zeno regime \cite{EARV03}, because the \qc
memory state is partially rectified even when the measurement gives the
expected result. The Zeno effect is known to be linked to error correction
from the very early days of quantum computation \cite{zeno_and_qc}; it has
been suggested as a stabilisation strategy by Shor \cite{Shor97} in his
famous paper on prime factoring, in a context identical to the current one
(i.e. syndrome measurement without a recovery circuit).

Of course, a direct observation of a fidelity jump, like that in figure
\ref{fig:nsygat-anomalous}, is not possible on a real \qc. It is true that
the result of each ancilla measurement is accessible, so one could decide
to purge a set of real experiments of those instances which showed a
``faulty'' reinitialisation; another option could be not to reinitialise
the ancilla at all. In any case, the only meaningful quantity is the
average fidelity.

\begin{figure}[tp]
  \myfig[(a)]{.8}{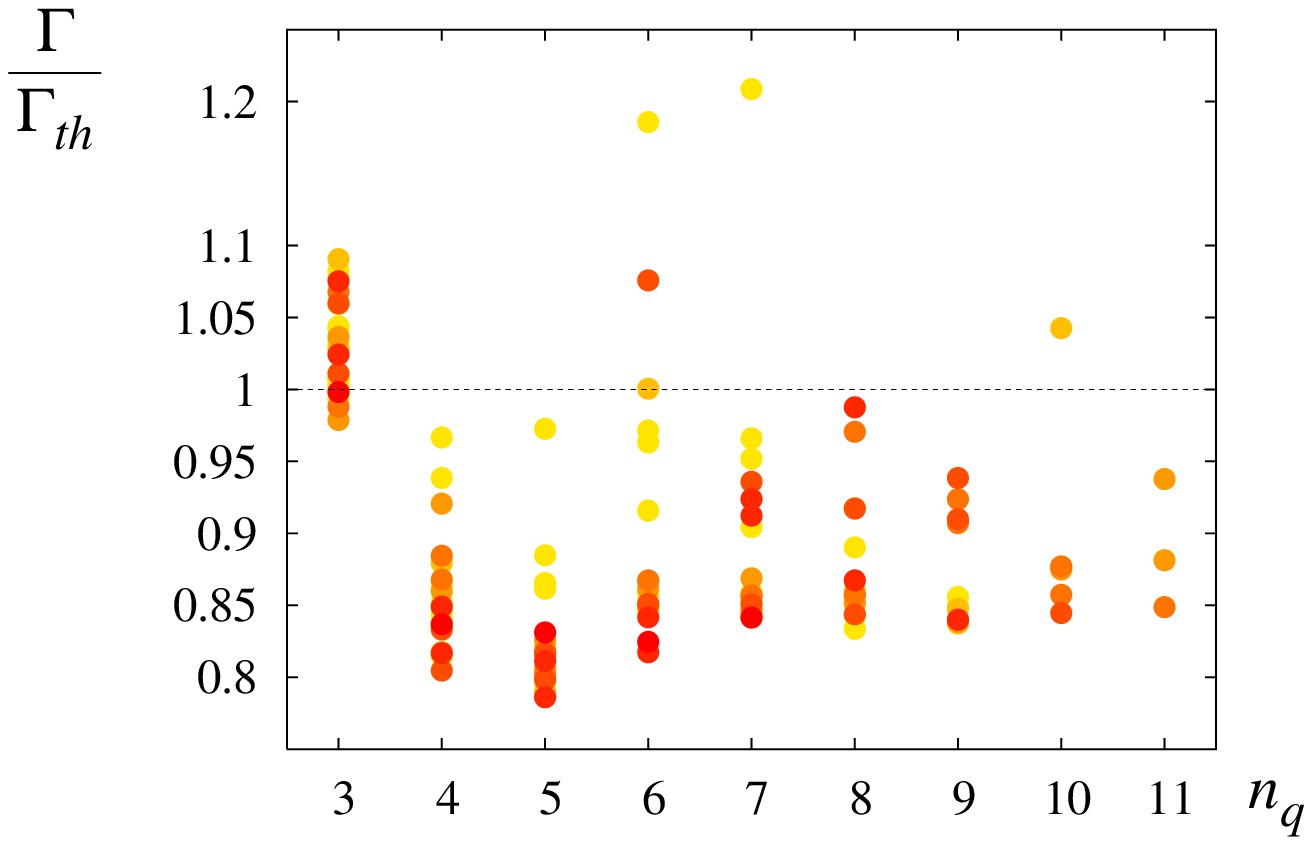}
  \myfig[(b)]{.8}{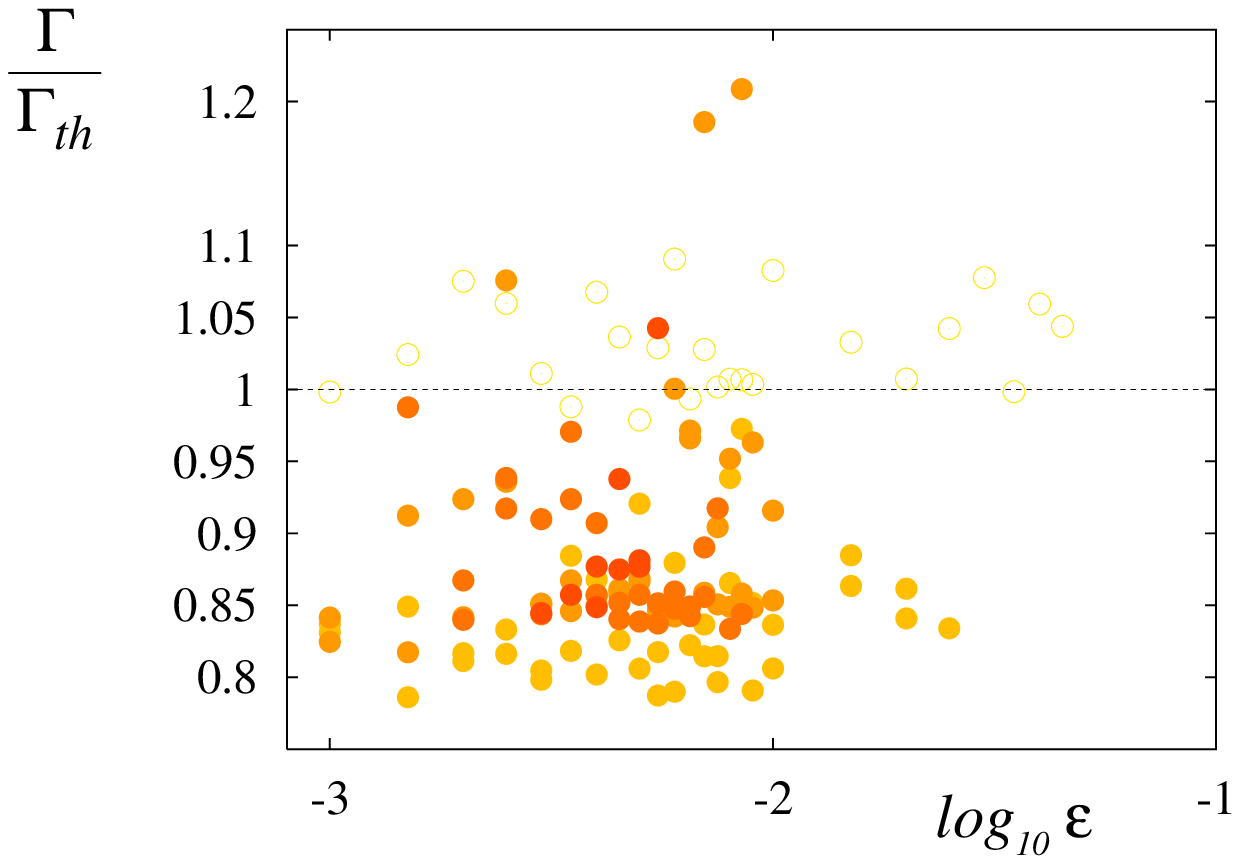}
  \myfig[(c)]{.8}{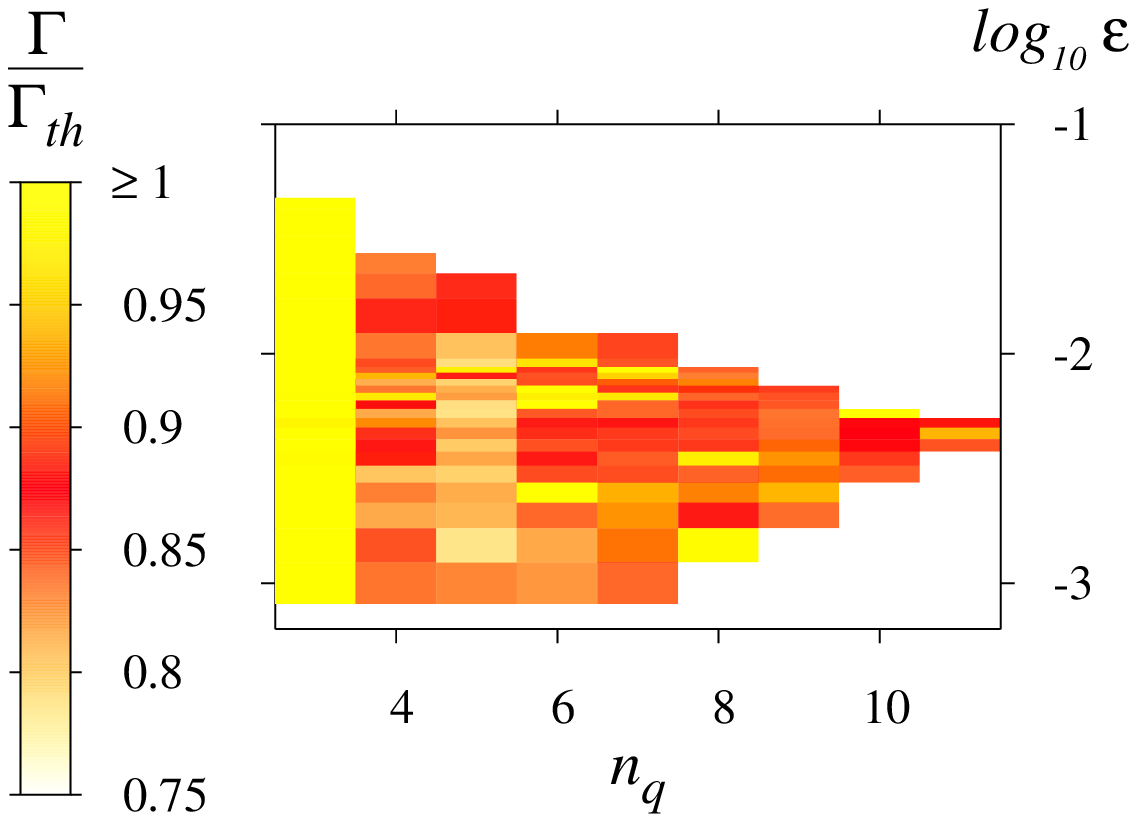}
  \caption[Fidelity degradation in the double-well map]{
    \label{fig:nsygat-gamma} \conline \\
    {\bf(a)} Ratio $\Gamma/\Gamma_{th}$ versus number of qubits $n_q$ for
    the double-well map. $\Gamma$ is the (numerical) fidelity decay
    constant and $\Gamma_{th}$ is its theoretical average, given by formula
    \ref{eq:fidelity-mean}. Each point corresponds to an average at fixed
    $n_q$ and $\epsilon$ with $200$ numerical simulations; at least $20$
    map iterations can be simulated before $f$ drops below $0.9$. Yellow
    (light) points correspond to large errors ($\epsilon \sim 10^{-1}$)
    while red (dark) points to small errors ($\epsilon \sim 10^{-3}$). For
    $n_q=3$ the algorithm does not use ancillae. \\
    {\bf(b)} Same as picture \ref{fig:nsygat-gamma}-a, but the independent
    variable is the error intensity $\epsilon$, while the yellow to red
    (light to dark) transition corresponds to an increasing number of
    qubits. Most of the points for $n_q \ge 4$ (filled circles) lie in the
    $80\%-90\%$ region, with a preference for $\sim 85\%$. Empty circles
    correspond to the case $n_q=3$ (no ancilla). \\
    {\bf(c)} Same as picture \ref{fig:nsygat-gamma}-a/b, but the dependence
    on the error intensity $\epsilon$ and the number of qubits $n_q$ is
    shown at the same time.  Each coloured box corresponds to an average at
    fixed $n_q$ and $\epsilon$ with $200$ numerical simulations. The orange
    and red (dark) boxes correspond to a $\Gamma$ which is $\sim 15\%$
    smaller than its theoretical value due to the reinitialisations of the
    ancilla (Zeno effect). For $n_q=3$ the algorithm implementation does
    not use ancillae.  Missing boxes are due either to combinations taking
    too long to simulate or to fidelities dropping too fast. \vspace{-3ex}
  }
\end{figure}

The numerical simulations presented in this article are not purged of the
unwanted reinitialisations. This slow-down is illustrated in figure
\ref{fig:nsygat-gamma}, where the numerically found value of $\Gamma$ is
approximately $10\%$ to $20\%$ smaller than the prediction of formula
\ref{eq:fidelity-mean}, with a preference for $\sim 15\%$. Note that for
$n_q = 3$ the algorithm implementation does not use ancillae, and the
result is in agreement with what was found for the sawtooth map (figure
\ref{fig:nsygat-gamma-st}). ``Wrong measurement'' events, like that shown
in figure \ref{fig:nsygat-anomalous}, become more and more frequent in the
averaging ensemble for large error intensities.

\section{Conclusions}

In this article the degradation of the performance of a \qc affected by
\nsygat is investigated, both analytically and numerically, by studying the
fidelity of the noisy states with respect to the ideal state (definition
\ref{eq:fidelity}): the average value of the fidelity after a fixed number
of elementary gates characterises the ``precision'' of the \qc and can be
efficiently extracted from a set of experiments via forward-backward
evolution. The numerical tests are performed using two algorithms that
simulate the sawtooth and the double-well maps. In the following $n_q$ is
the number of qubits in the \qc and $N=2^{n_q}$ is the number of states.

The degradation of the fidelity is shown to be determined basically by
$\sigma_\err^2$, the mean variance of the phases of the eigenvalues of the
error operators (definition \ref{eq:sigmastar}); stated in this way, this
result is valid for every choice of elementary gates. In a realistic case,
where the \qc is endowed with a particular set of elementary operations,
each characterised by a specific error operator distribution, this quantity
is algorithm-dependent only through the relative frequencies of the
elementary gates; so, $\sigma_\err^2$ reflects the effective decoherence
inherent to the \qc, and a careful analysis of the results of different
algorithms would allow in principle to determine the proper $\sigma_\err^2$
for each gate type.

The noise-averaged fidelity after $N_g$ elementary gates is found to be
$\mean{f_{N_g}} = 1 - \frac{N}{1+N} N_g \sigma^2_\err + O(\sigma^3_\err)$.
The fluctuations of the value of the fidelity after a fixed number of gates
are such that $\sigma(f)/(1-\mean{f})$ depends only on $n_q$ and is
exponentially small with it. It is also found that the reinitialisation of
the auxiliary qubit during the computation for the double-well map reduces
the fidelity decay rate of $\sim 15\%$ (this is known in literature as an
error correction in the Zeno regime).

Note that a particular case of distribution of the error operators is the
one which does not depend on the gate type (this case bears the additional
bonus that there is no dependence at all on the algorithm). In this
scenario, an ensemble of unitary errors is equivalent to the more commonly
studied case of a non-unitary error channel, like for instance the
depolarisation channel. 
%The only assumption in the derivation of the law for
%$\mean{f_{N_g}}$ is the absence of correlation among errors at different
%steps.

The rather good agreement between the quantitative predictions of the
theory and the numerical results for the double-well and sawtooth maps
suggests a general validity for the assumption on the distribution of the
states visited by an algorithm. It is clear however that there exist
trivial algorithms which do not follow the predicted behaviour for some
choice of their initial state. The question is then to characterise the
class of algorithms which satisfy well enough the predictions, and, in
particular, to understand whether this is linked or not to the fact that
the algorithms used as examples exploit the \qc in order to simulate
another quantum system, showing ergodic properties. Due to the robustness
of the prediction for $\sigma_\err^2$ for state distributions close to the
uniform distribution, the conjecture is that the class is not limited to
this type of algorithms.

It would be interesting to extend the ideas presented in this article in
order to include measurements of auxiliary qubits; as shown by numerical
simulations, measurements of qubits supposed to be, in absence of errors,
in a known state can slow down the fidelity decrease. In perspective, this
could allow to deal with error correction codes, a subject which has not
been treated here.

The numerical simulations in this article were performed using a freely
available implementation of the quantum programming language described in
\cite{BSC03}. The author acknowledges useful discussions with A. Pomeransky,
M. Terraneo, B. Georgeot and J. Emerson.

This work was supported in part by the European Union {\small RTN} contract
{\small HPRN-CT-2000-0156} ({\small QTRANS}) and {\small IST-FET} project
{\small EDIQIP}, and by the French government ACI (Action Concert\'ee
Incitative) Nanosciences-Nanotechnologies {\small LOGIQUANT}.

\appendix

\section{Hamiltonian maps as quantum algorithms} \label{app:algorithms}

The goal of this appendix is to show the general approach to build a
quantum algorithm for a quantum Hamiltonian map derived from a classical
unidimensional kicked map. A classical kicked map is the stroboscopic
observation of the phase space of a system which is affected by an
impulsive conservative force given by a periodic potential at regular
intervals in time and which evolves freely in the meanwhile. If the
observation occurs just before the kick, the map connecting the points in
the dimensionless $(\theta,\eta)$ phase space can be written, without loss
of generality, as
\begin{equation} \label{eq:map}
  \left\{ \begin{array}{lll}
    \bar{\eta} & = & \eta - k \frac{dV}{d\theta} \\
    \bar{\theta} & = & \theta + T\bar{\eta} \pmod{2\pi}
  \end{array} \right.
\end{equation}

In this expression, $T$ represents the (dimensionless) time interval between
two kicks, $2\pi$ is the period of the potential $V(\theta)$ and $k$ is a
parameter which governs the intensity of the potential.  It is easy to show
that, if the rescaled variable $\eta'=T\eta$ is used instead of $\eta$, the
map depends on one parameter only, that is $K=kT$. In the map \ref{eq:map}
the $\theta$ range is finite, while the momentum $\eta$ can vary
arbitrarily. However, the transformation
\begin{equation} \label{eq:equivEta}
  (\theta,\eta) \rightarrow (\theta,\eta+2\pi L/T), \quad\quad L\in
  \mathbb{Z},
\end{equation}
is a symmetry for the system; this invariance identifies a natural cell,
with an extension of $2\pi$ in both directions of the phase space, and $L$
can therefore be associated to the number of cells.  The time-dependent
classical Hamiltonian driving the evolution described by the map
\ref{eq:map} is
\begin{displaymath}
  H(\theta,\eta,t) = \frac{\eta^2}{2} +
      kV(\theta)\sum_{j=-\infty}^{+\infty} \delta(t-jT).
\end{displaymath}

The map quantisation can be accomplished by replacing the dynamical
variables $\theta$ and $\eta$ with the corresponding operators
$\hat{\theta}$ and $\hat{\eta}$. In the $\theta$ representation, the
momentum operator becomes, as usual, $\hat{\eta}=-i\frac{\partial}
{\partial\theta}$. The quantum equivalent of the map \ref{eq:map} is the
evolution operator calculated for a time interval $T$ (which includes
always one kick), known as the Floquet operator ($U_F$).  Given that the
action of the potential is impulsive, the integration of the quantum
Hamiltonian is very easy:
\begin{equation} \label{eq:opFloquet}
  \hat{U}_F = \exp\left(-i\frac{\hat{\eta}^2}{2\hbar}T\right)
       \exp\left(-i\frac{kV(\hat{\theta})}{\hbar}\right).
\end{equation}

Therefore, in the quantum case the dynamics is determined not only by the
product $K=kT$, but also by the ``magnitude'' of $\hbar$ (for instance, by
the ratio $k/\hbar$).

The simulation of the quantum map on a finite \qc implies some type of
discretisation, which introduces an additional parameter, namely the number
of available levels, $N=2^{n_q}$.  The easiest choice is to associate the
computational basis states \ket{i} to some eigenvalue $\theta_i$. In this
way the ``position'' operator $\theta$ becomes diagonal in the
computational basis. In the algorithm described in this article the
following choice holds:
\begin{equation} \label{eq:choiceTheta}
  \ket{i}\quad\longrightarrow\quad\theta_i=\frac{2\pi}{N}(i+\delta_{\theta}).
\end{equation}

In this way the position eigenvalues are equidistant; $\delta_{\theta}$ is
an offset which fixes the exact $2\pi$ interval where the eigenvalues are
located. The choice $\delta_{\theta}=(1-N)/2$ corresponds to
$-\pi<\theta_i<\pi$, with perfect symmetry around $\theta=0$.  The
periodicity condition $\theta \equiv \theta+2\pi$ implies that the
eigenvalues of the momentum must be quantised in units of $\hbar$.  The
finiteness of the \qc memory allows only for a finite number $N$ of
eigenvalues of the momentum to be simulated. It is therefore natural to
resort to a cyclic condition, that is assuming that the range of the
eigenvalues corresponds to an integer number of cells (see the
transformation \ref{eq:equivEta}):
\begin{equation} \label{eq:Th}
  \Delta\eta = N\hbar = 2\pi L/T
  \quad\quad\Rightarrow\quad\quad
  T\hbar = \frac{2\pi L}{N}.
\end{equation}

The last step in the construction of the algorithm is the fact that the
quantum Fourier transform $\mathcal{F}$ ``exchanges'' the momentum
representation with the position representation. This means that if, at
some time during the simulation, the \qc memory's state is $\ket{\psi} =
\sum_j a_j\ket{j}$ and it is associated to the simulated system's state
$\sum_j a_j\ket{\theta_j}$, where the \ket{\theta_j}'s are the eigenstates
of $\hat{\theta}$ with eigenvalue $\theta_j$, then
$\mathcal{F}^\dagger\ket{\psi} = \sum_j b_j\ket{j}$ must be associated to
the simulated system's state $\sum_j b_j\ket{\eta_j}$, where the
\ket{\eta_j}'s are the eigenstates of $\hat{\eta}$ with eigenvalue $\eta_j
= j\hbar$. In other words, the Floquet operator $\hat{U}_F$ (see
\ref{eq:opFloquet}) can be simulated by a quantum circuit $Q$ whose action is:
\begin{eqnarray}
  \label{eq:opFloqCirc} Q &=&
     \mathcal{F}\circ Q_{\eta}\circ\mathcal{F}^{\dagger}\circ Q_{\theta} \\
  \label{eq:opEtaCirc} Q_{\eta}\ket{j} &=&
     \exp\left(-i\frac{\eta_j^2}{2\hbar}T\right)\ket{j} \\
  \label{eq:opThetaCirc} Q_{\theta}\ket{j} &=&
     \exp\left(-i\frac{kV(\theta_j)}{\hbar}\right)\ket{j} \\
  \label{eq:opFourier} \mathcal{F}\ket{j} &=&
     \frac{1}{\sqrt{N}}\sum_m e^{2\pi i\frac{jm}{N}}\ket{m}
\end{eqnarray}

\noindent It is now useful to define an operator $\expower{\beta}{p}$ as
\begin{equation} \label{eq:defexpower}
  \expower{\beta}{p}\ket{j} = \exp\left(2\pi i\beta j^p\right)\ket{j}.
\end{equation}

The circuit $Q_{\eta}$ acting in the momentum representation (see equation
\ref{eq:opEtaCirc}) can then be simplified by using the definition
\ref{eq:defexpower}, and replacing the eigenvalue $\eta_j$ with $j\hbar$
and the product $T\hbar$ with the help of expression \ref{eq:Th}:
\begin{equation}  \label{eq:circuitQeta}
  Q_{\eta}\ket{j} = \exp\left(-2\pi i\frac{L}{2N}j^2\right)\ket{j} =
  \expower{\frac{-L}{2N}}{2}\ket{j}.
\end{equation}

In this expression the parameters $T$ and $\hbar$ have been replaced by the
number of levels $N$ and the number of cells $L$: the free evolution is
thus completely geometrical. The (dimensionless) quantum action $\hbar$ is
proportional to $L/N$, therefore the semi-classical limit can be attained by
taking $L/N \rightarrow 0$. \par The circuit $Q_{\theta}$ (see equation
\ref{eq:opThetaCirc}) is easy to manage when the potential $V(\theta)$ of
the map \ref{eq:map} is a polynomial in $\theta$ (let $p$ be its degree).
One can in fact introduce the polynomial $\mathcal{P} (x + \delta_\theta) =
\sum_{m=0}^p f_m x^m$ and set

\begin{displaymath}
  V(\theta_j) = V\left(\frac{2\pi}{N}(j+\delta_\theta)\right) =
  \left(\frac{2\pi}{N}\right)^p \mathcal{P}(j+\delta_\theta),
\end{displaymath}
where $\theta_j$ has been replaced with expression \ref{eq:choiceTheta}.
The circuit $Q_{\theta}$ (see equation \ref{eq:opThetaCirc}) can then be
broken down into circuits of type \expower{\beta}{m} (see the definitions
\ref{eq:defexpower} and \ref{eq:Th} and remember that $K=kT$):
\begin{eqnarray} \label{eq:circuitQtheta}
  Q_{\theta}\ket{j} &=& \exp\left(-i\frac{KV(\theta_j)}{2\pi
     L/N}\right)\ket{j} \nonumber \\ &=& \prod_{m=1}^p
     \expower{\frac{-Kf_m(2\pi)^{p-2}}{LN^{p-1}}}{m}\ket{j}. \qquad
\end{eqnarray}

Note that the term $m=0$, which gives only a global phase, can be
neglected.  For the double-well \cite{ChepelianskiiShepelyansky02} map
($p=4$) the polynomial is:
\begin{displaymath}
  \mathcal{P}(y)=\left(y^2-a^2\right)^2,
\end{displaymath} 
where $a$ is an additional parameter which locates the centres of the two
wells. The coefficients $f_m$ are therefore
\begin{displaymath}
  f_4=1, \quad f_3=4\delta_\theta, \quad f_2=6\delta_\theta^2-2a^2, \quad
  f_1=4\delta_\theta(\delta_\theta^2-a^2).
\end{displaymath}

For the sawtooth \cite{BCMS01} map ($p=2$) the polynomial and the
coefficients are respectively
\begin{displaymath}
  \mathcal{P}(y)=-\frac{1}{2}y^2, \quad f_2=-\frac{1}{2}, \quad
  f_1=-\delta_\theta.
\end{displaymath} 

Summarising, it has been shown that the circuit $Q$ (equation
\ref{eq:opFloqCirc}) can be written as a composition of circuits for the
Fourier transform $F$ (equation \ref{eq:opFourier}) and for the operators
$\expower{\beta}{m}$ (equation \ref{eq:defexpower}).  The parameters which
govern the simulation are the classical parameter $K$ (integrability
$\rightarrow$ chaos), the number of simulated cells $L$ and the number of
available levels $N$ ($L/N\rightarrow 0$ being the semi-classical limit) in
addition, of course, to the functional form of the potential $V(\theta)$.

The state of the \qc memory after each application of the circuit $Q$
corresponds to the coordinate representation for the simulated system, but
trivial changes allow the simulation in the momentum representation.  It is
well known that the circuit for $\mathcal{F}$ can be implemented
efficiently \cite{Coppersmith02}. Appendix \ref{app:expower} shows that
also $\expower{\beta}{m}$ can be implemented efficiently; the conclusion is
therefore that the circuit for the simulation of a quantum kicked
Hamiltonian with a polynomial potential can be implemented efficiently on a
\qc.

There exist, of course, interesting potentials which are not polynomial;
for instance, the kicked rotator map has $V(\theta) = \cos \theta$. For
this case, two approaches are known. In \cite{GeorgeotShepelyansky00} an
auxiliary register and a number of additional ancillae is used for
calculating $\ket{\theta}\ket{0} \rightarrow \ket{\theta}\ket{\cos\theta}$
with a finite number of digits in the mantissa; the auxiliary register is
subsequently used to implement $Q_\theta$. If the size of the mantissa for
the calculation of the cosine is $O(n_q)$, the circuit depth is $O(n_q^3)$
and the number of ancillae is $O(n_q)$. In \cite{PomeranskyShepelyansky03}
another approximated method, with circuit depth $O(kn_q)$, is introduced.
This method is very interesting because no auxiliary qubit is required;
however, the precision with which $Q_\theta$ is implemented is $O(k)$, so
the approach is optimal only for small $k$ values.

\section{A quantum circuit for exponentiation} \label{app:expower}

The goal of this appendix is to show a procedure for building an efficient
quantum circuit implementing the unitary transformation
\begin{equation} \label{eq:expower}
  \expower{\beta}{p}\ket{x} = e^{2\pi i\beta x^p}\ket{x}.
\end{equation}

In expression \ref{eq:expower}, the exponent $p$ is a positive integer
while $\beta$ is a real coefficient. \ket{x} is the $x$-th element of the
computational basis, therefore $\expower{\beta}{p}$ is diagonal in this
basis. The circuit is implicitly parametrised by the number $n_q$ of qubits
in the register which $\expower{\beta}{p}$ operates on. The first step is
to write the integer $x$ labelling \ket{x} as a binary string, $x =
\sum_{j=0}^{n_q-1} a_j 2^j$. By replacing this expression into the phase of
definition \ref{eq:expower}, it is possible to rewrite it as the product:
\begin{displaymath}
  e^{2\pi i\beta x^p} = \prod_{j_1 \dots j_p} e^{2\pi i\beta a_{j_1}\dots
  a_{j_p} 2^{j_1+\dots+j_p}}.
\end{displaymath}

Since the coefficients $a_j$ are binary digits, the product $a_{j_1}\dots
a_{j_p}$ is zero (i.e. the phase is trivial) unless all the $a_j$'s are
equal to $1$. Therefore each factor in the previous expression corresponds
to a multi-controlled phase gate in the circuit for $\expower{\beta}{p}$
of definition \ref{eq:expower}. This gate acts on the qubits selected by
the set $W(j_1\dots j_p)$, which contains the values of the indexes $j_1$,
\dots $j_p$ with neither repetitions nor order. Introducing the notation
$\mathcal{C}_W(\varphi)$ for a multi-controlled phase gate applying the
phase $\exp(2\pi i\varphi)$ to the qubits in the set $W$, one obtains
\begin{equation} \label{eq:multicontrolled}
  \expower{\beta}{p} = \prod_{j_1 \dots j_p} \mathcal{C}_{W(j_1\dots
  j_p)}(\beta 2^{j_1+\dots+j_p}).
\end{equation}

Up to now, this method follows exactly the procedure described in
\cite{ChepelianskiiShepelyansky02}. The set $W$ does not contain
duplicated elements, so that its cardinality is smaller or equal to $p$
(but it is always positive).  Since neither the sum $j_1+\dots+j_p$ nor the
set $W(j_1\dots j_p)$ depend on the order of the indexes, all the gates
concerning the same indexes and differing only in their order can be
collected into a single gate. Let
\begin{equation} \label{eq:defP}
  P = \left\{ J \subset \{0,...,n_q-1\} | J\ne\emptyset \wedge \card{J}
    \leq p \right\},
\end{equation}
where $0 \dots n_q-1$ are the possible values for the qubit indexes. 
Then expression \ref{eq:multicontrolled} can be rewritten as
\begin{eqnarray} \label{eq:U2}
  \expower{\beta}{p}
        &=& \prod_{J \in P} \prod_{\substack{j_1\dots j_p\\W(j_1\dots j_p)=J}}
        \mathcal{C}_{W(j_1\dots j_p)}(\beta 2^{j_1+\dots+j_p}) \nonumber \\
        &=& \prod_{J \in P} \mathcal{C}_J\Big(\beta \hspace{-3.5ex}
        \sum_{\substack{j_1\dots j_p \\ W(j_1\dots j_p)=J}}
        \hspace{-3.5ex} 2^{j_1+\dots+j_p}\Big),
\end{eqnarray}
where in the last line the gates with the same $J$ have been compressed
into a single gate. Note that two gates act on the same qubits if and only
if their $W$ sets are equal. To proceed, one needs to define the
set $G_k(n)$ of the partitions of $n$ objects in exactly $k$ non-empty
subsets. A partition $g \in G_k(n)$ is then a sequence of $k$ positive
integers $\{g_1 \dots g_k\}$ such that their sum is equal to $n$. Given a
set $J$ and a partition $g \in G_{\card{J}}(p)$, the pair $\langle J,g
\rangle$ corresponds to a subdivision of the set of $p$-tuples whose $W$
set is $J$. It is then possible to apply the replacement
\begin{equation} \label{eq:replace}
  \sum_{\substack{j_1\dots j_p\\W(j_1\dots j_p)=J}}
  \quad\longrightarrow\quad\quad \sum_{g\in G_{\card{J}}(p)}
  \sum_{\substack{j_1\dots j_p \\ W(j_1\dots j_p)=J \\ j_1+\dots+j_p = \sum
  g_i J_i}}
\end{equation}

For each pair $\langle J,g \rangle$, the sum $j_1+\dots+j_p$ is fixed,
therefore the last summation on the right of expression \ref{eq:replace}
can be replaced by its multinomial weight $\frac{p!}{g_1! \cdots
g_{\card{J}}!}$. Putting all together, one finally obtains that the
operator $\expower{\beta}{p}$ can be replaced by a product of
multi-controlled gates $\mathcal{C}_J(\varphi)$, with a bijective
correspondence between the gates and the sets $J\in P$, where
\begin{equation} \label{eq:multifinal}
   \varphi = \beta\sum_{g\in G_{\card{J}}(p)} 
        \frac{p!}{g_1! \cdots g_{\card{J}}!} \, 2^{\sum g_i J_i}.
\end{equation}

This gate collection method presents the obvious advantage that the circuit
depth decreases. In fact, if one gate is built for each possible index
combination $j_1\dots j_p$, the total number of gates is trivially
$(n_q)^p$. If, on the other hand, the previously described compression
is applied, one obtains as many gates as the cardinality of the set of
partitions $P$ (definition \ref{eq:defP}), i.e.
\begin{equation} \label{eq:cardin}
  \card{P} = \sum_{k=1}^{\min(n_q,p)} \binom{n_q}{k}.
\end{equation}

For $n_q\leq p$ this sums to $2^{n_q} - 1$. However, it is more interesting
to consider the limit $n_q \gg p$; approximating the sum with the help of
Stirling's formula one obtains
\begin{displaymath}
  \card{P} \sim \frac{(n_q)^p}{p!}.
\end{displaymath}

It is easy to check that this approximation is already good for $n_q=5$ and
$p=4$, where the savings $1/p!$ correspond to $\sim 95\%$ of the gates. It
should be kept in mind that the calculation of $\varphi$ in expression
\ref{eq:multifinal} can be numerically critical, because for large $n_q$
the integer multiplying $\beta$ grows exponentially as $2^{p n_q}$. Since
this phase term is to be taken modulo 1, the least significant bits of the
representation of $\beta$ are to be handled carefully\footnote{This problem
of course does not originate from the collection method. Factors of order
$2^{pn_q}$ are present also in the ``plain'' approach.}.

In practice however, when $\expower{\beta}{p}$ is used as a block of a
larger circuit, like \ref{eq:circuitQeta} or \ref{eq:circuitQtheta},
$\beta$ is also exponentially small. For $p \geq 2$ the worst case comes
from circuit \ref{eq:circuitQeta}, where $\beta = O(2^{-n_q})$, so that the
phase $\varphi$ is $O(2^{n_q (p-1)})$. If $\beta$ is stored as a floating
point number, one loses $p-1$ bit of the mantissa of $\varphi$ for every
qubit added. For a double precision floating point\footnote{Take for
  instance the $64$ bit IEEE floating point format, which has $53$ binary
  digits available for the mantissa.} this problems arises around
$n_{\mathrm{crit}} = 18$ for the double-well map and around
$n_{\mathrm{crit}} = 53$ for the sawtooth map. This is beyond or at the
limit of the possibilities of current \qc simulators. In any case, the
study of the real degradation of a computation due to this loss of
precision in the most critical gates for $n_q > n_{\mathrm{crit}}$ is not
trivial and has not been performed yet.

\section{Error models and bounds to the fidelity degradation}
\label{app:error_model_nsygat}

In all the numerical simulations of this article, it is supposed that the
\qc is endowed with the following set of hardware operations:
\begin{displaymath}
  H = \frac{1}{\sqrt{2}} \left(\begin{smallmatrix} 1&\phantom{-}1\\1&-1
      \end{smallmatrix}\right) \quad R_\phi = \left(\begin{smallmatrix}
      1&\\&\!e^{i\phi}\!\! \end{smallmatrix}\right) \quad CR_\phi =
      \left(\begin{smallmatrix} 1&&&\\&1&&\\&&1&\\&&&\!e^{i\phi}\!\!
      \end{smallmatrix}\right),
\end{displaymath}
where $\phi\in\mathbb{R}$ is a phase (indeed, it can be only a multiple of
some ``atomic'' phase, like $2\pi/2^{32}$, which mimics the limited
precision of the classical control system). $H$ is the Hadamard
transformation, while $R_\phi$ and $CR_\phi$ are phase shifts and
controlled phase shifts respectively. In general, the controlled operation
$CA$ corresponds to the matrix $\left(\begin{smallmatrix} \id& \\ &A
\end{smallmatrix}\right)$. It is sometimes useful to express the
elementary gates as rotations $R_{\hat{\eta}}(\alpha)$ of an angle $\alpha$
around an axis $\hat{\eta}$:
\begin{displaymath}
  R_{\hat{\eta}}(\alpha) =
  e^{-i\frac{\alpha}{2}\hat{\eta}\cdot\vec{\sigma}} =
  \cos\left(\frac{\alpha}{2}\right) - i\sin\left(\frac{\alpha}{2}\right)
  \hat{\eta}\cdot\vec\sigma.
\end{displaymath}

With this notation, setting $\widehat{xz}=(\hat{x}+\hat{z})/\sqrt{2}$, one
finds $H \!= e^{i\frac{\pi}{2}} R_{\widehat{xz}}(\pi) \equiv
R_{\widehat{xz}}(\pi)$ and $R_\phi = e^{i\frac{\phi}{2}} R_{\hat{z}}(\phi)
\equiv R_{\hat{z}}(\phi)$, where '$\equiv$' means equivalence modulo a
global phase (which can be neglected). Note, however, that $CR_\phi =
e^{i\frac{\phi}{4}} [R_{\hat z} (\frac{\phi}{2})\otimes\id] \circ CR_{\hat
z}(\phi) \not\equiv CR_{\hat z} (\phi)$, so that $CR_\phi$ and $CR_{\hat
z}(\phi)$ are inequivalent even modulo a global phase.

In the \nsygat error model (see for instance \cite{SongShepelyansky01}),
each gate is replaced by a still unitary transformation, close in norm to
the original gate, parametrised by a single error variate $\xi$ with flat
probability in $[-\epsilon/2, \epsilon/2]$. The quantity $\epsilon$ is
called the ``error intensity'', and it summarises the amount of noise which
affects the \qc. It is useful to express a perturbed elementary gate as a
composition of the unperturbed gate followed by an error operator close to
the identity. The imperfections for $z$-rotations and controlled
$z$-rotations are implemented by shifting their phase angle by an amount
$\xi$; it is easy to show that
\begin{eqnarray*}
  \textrm{noisy~}R_\phi &\longrightarrow& R_\xi R_\phi
  \equiv R_{\hat{z}}(\xi) R_\phi \qquad \mathrm{and}\\
  \textrm{noisy~}CR_\phi &\longrightarrow& CR_\xi CR_\phi
  \equiv \left[R_{\hat z}(\frac{\xi}{2}) \otimes \id\right] 
  CR_{\hat z}(\xi) \, CR_\phi.
\end{eqnarray*}

Each Hadamard gate is transformed into a rotation of an angle $\pi$ around
an axis $\hat\eta'$ randomly tilted around the unperturbed direction
$\hat\eta = \widehat{xz}$ of an angle\footnote{Contrary to previous works
\cite{SongShepelyansky01}, the tilting angle for $H$ is $\xi/2$ and not $\xi$.
This uniforms the effects of \unoq gates, because all \unoq error operators
are rotations of an angle $\xi$ around some axis.} $\xi/2$. The perturbed
gates can be generated by taking a random direction $\hat\mu$ in the plane
orthogonal to $\widehat{xz}$ and setting $\hat\eta' = \hat\eta
\cos\frac{\xi}{2} + \hat\mu\times\hat\eta \sin\frac{\xi}{2}$. Note that the
signs of $\mu$ and $\xi$ are not independent. A little algebra shows that
\begin{displaymath}
  \textrm{noisy~}H \longrightarrow R_{\hat{\mu}}(\xi) H.
\end{displaymath}

Thus, the error operators $\opE_k$ in the \nsygat error model are
$R_{\hat{z}}(\xi)$, $R_{\hat{\mu}}(\xi)$ and $CR_\xi$, the phases of the
eigenvalues for \unoq errors are $\{\pm \frac{\xi}{2}\}$, and those for
\dueq errors are $\{0_3,\xi\}$. All these eigenvalues are of course
specified modulo a global phase shift. It is easy to see that this model is
unbiased. Another approach to \nsygat \cite{GeorgeotShepelyansky01}
consists in diagonalising the gates and perturbing the eigenvalues of an
amount $\xi$; given that the average properties of the induced effective
decoherence are determined only by the error operators' spectrum, it is not
surprising that this alternative approach yields in general very similar
results.  The calculation of the radius $r_k$ of the neighbourhood of the
eigenvalue phases (see equation \ref{eq:intro_rk}) is very simple; $r_k$ is
always $|\xi|/2$, so that (see equation \ref{eq:defvarsigmaerr})
\begin{equation} \label{eq:varsigma_nsygat}
  \varsigma_{\err} \simeq \left\langle r_k \right\rangle 
  = \frac{1}{\epsilon} \int_{\frac{\epsilon}{2}}^{\frac{\epsilon}{2}}
    \frac{|\xi|}{2} \,d\xi = \frac{\epsilon}{8}.
\end{equation}

The incoherent replacement $(\sum_k \varsigma_{\err k})^2 \rightarrow
(\sum_k \varsigma_{\err k}^2)$ in formula \ref{eq:boundU} gives the result
necessary for formula \ref{eq:fidelity4}:
\begin{equation} \label{eq:varsigma_incoherent}
  1-f \leq N_g \left\langle \frac{\xi^2}{4} \right\rangle
  = \frac{N_g}{\epsilon} \int_{\frac{\epsilon}{2}}^{\frac{\epsilon}{2}}
    \frac{\xi^2}{4} \,d\xi = \frac{\epsilon^2 N_g}{48}
\end{equation}

\section{Uniform averages and concentration of measure} \label{app:averages}

This appendix details the evaluation of an integral like
\begin{equation} \label{eq:genaverage}
   \mathcal{I}(\opR\apc{1},\dots,\opR\apc{r}) =
   \int  d\mu(\psi) \,\prod_{k=1}^r \bra{\psi}\opR\apc{k}\ket{\psi},
\end{equation}
where the $\opR\apc{k}$ are operators over an $N$-dimensional Hilbert
space, and $d\mu(\psi)$ is the uniform measure over its unity vectors (that
is, the unique invariant measure under the action of all unitary matrices).
If the integrand is expanded over some basis, then
\begin{displaymath}
  \mathcal{I}(\opR\apc{1},\dots,\opR\apc{r}) = \prod_k R\apc{k}_{i_kj_k}
  \cdot \underbrace{\int d\mu(\psi) \prod_k \psi^*_{i_k}\psi_{j_k}}_
  {T_{i_1j_1\dots i_rj_r}} \quad,
\end{displaymath}
where the linearity of the expectation value and the fact that the average
is over the $\ket{\psi}$'s and not over the $\opR\apc{k}$'s were used
(summation over repeated indexes is understood). The measure $d\mu(\psi)$
over the Hilbert space can be turned into a measure $d\mu(U)$ over the set
of unitary transformations of this space. Since $d\mu(\psi)$ is by
definition invariant when $U\ket{\psi}$ is substituted for $\ket{\psi}$ in
formula \ref{eq:genaverage}, with $U$ a generic unitary matrix, $d\mu(U)$
remain completely determined (Haar or uniform measure, corresponding to the
unitary circular ensemble \cite{Mehta91}).

This invariance allows for the calculation of the averages
\ref{eq:genaverage} without a direct integration \cite{unitary_integrals}.
It can be shown \cite{Brouwer97} that the tensor $T_{i_1j_1\dots i_rj_r}$
must be proportional to a sum of products of Kronecker's deltas symmetric
with respect to the indexes of the same type (i.e. the $i$'s or the $j$'s);
the normalisation coefficient can then be fixed by comparison with the case
$\opR\apc{k} = \id \,\,\forall k$. For $r=1$
\begin{equation} \label{eq:average1}
  T_{ij} \propto \delta_{ij}, \qquad \mathrm{therefore} \qquad
  \mathcal{I}(\opR) = \frac{\tr(\opR)}{N}.
\end{equation}

\noindent For the case $r=2$ one similarly obtains
\begin{displaymath}
  T_{ijkl} \propto (\delta_{ij}\delta_{kl} + \delta_{il}\delta_{jk}).
\end{displaymath}

\noindent Thus, the integral \ref{eq:genaverage}
can be written using only traces:
\begin{equation} \label{eq:average2-general}
  \hspace{-1pt} R\apc{1}_{ij}R\apc{2}_{kl}T_{ijkl}
  = \frac{\tr(\opR\apc{1})\tr(\opR\apc{2})
        + \tr(\opR\apc{1}\opR\apc{2})}{N^2 + N}.
\end{equation}

If the operators are simultaneously diagonalisable, it is easy to see that
the integral depends only on their spectrum. If $\opR$ is a generic unitary
operator, setting $\mathcal{I}(\opR,\opR^\dagger) = \mathcal{I}_2(\opR)$
and $A=\frac{N}{1+N}$ one obtains that
\begin{equation} \label{eq:average2}
  \mathcal{I}_2(\opR) = 1 - A \Big[ 1 - \frac{|\tr(\opR)|^2}{N^2} \Big].
\end{equation}

When $\opR$ has only two eigenvalues, $e^{i\lambda_1}$ and
$e^{i\lambda_2}$, with the same multiplicity $N/2$ (i.e., when it is a
generic \unoq operator), the previous formula reduces to
\begin{equation} \label{eq:average_unoq}
  \mathcal{I}_2\left(\begin{smallmatrix}
        e^{i\lambda_1}&0\\0&e^{i\lambda_2}\end{smallmatrix}\right)
  = 1 - A\sin^2 \left({\textstyle\frac{\lambda_1-\lambda_2}{2}}\right).
\end{equation}

It is interesting to calculate the same integral in a different way. By
introducing the probability $0\leq p\leq 1$ of being in the first
eigenspace, the state can be written as
\begin{displaymath}
  \ket{\psi} = \sqrt{p}\,\ket{\psi_1} + \sqrt{1-p}\,\ket{\psi_2},
\end{displaymath}
where the normalised state \ket{\psi_j} is an eigenstate with eigenvalue
$e^{i\lambda_j}$, and all the unnecessary phases have been reabsorbed. The
integral in formula \ref{eq:average_unoq} becomes then
\begin{eqnarray} \label{eq:average_unoq_alt}
  \hspace{-1cm}\mathcal{I}_2 \left(\begin{smallmatrix}
      e^{i\lambda_1}&0\\0&e^{i\lambda_2}\end{smallmatrix}\right)
  &=& \int_0^1\left|pe^{i\lambda_1} + (1-p)e^{i\lambda_2}\right|^2
  \mathcal{P}(p)dp \nonumber \\
  %%  && \hspace{-5em} = 1 - 4\sin^2\left({\textstyle
  %%        \frac{\lambda_1-\lambda_2}{2}}\right)
  %%        \int_0^1 p(1-p)\mathcal{P}(p)\,dp \\
  && \hspace{-5em} = 1 - 4\sin^2\left(
    {\textstyle \frac{\lambda_1-\lambda_2}{2}}\right)
  \cdot \left[ \mean{\,p\,} ( 1 - \mean{\,p\,} ) - \sigma^2_p \right].
\end{eqnarray}

A freedom in the choice of the probability density function $\mathcal{P}
(p)$ corresponds to a freedom for the measure $d\mu(\psi)$. Due to the
symmetry of $\lambda_1$ and $\lambda_2$ in the uniform distribution,
$\mean{\,p\,}$ is necessarily $1/2$. Formulae \ref{eq:average_unoq} and
\ref{eq:average_unoq_alt} then coincide when
\begin{displaymath}
  \sigma_p^2 = \frac{1-A}{4} = \frac{1}{4(1+N)}.
\end{displaymath}

In the framework of quantum computation, the dimension of the space is
exponential in the \qc size, $N=2^{n_q}$, so $\sigma_p^2$ is exponentially
small. This behaviour, known in literature as the {\em concentration of
  measure phenomenon} \cite{conc_measure}, is much more general than the
case studied here: in a space with a large number of dimensions, the
deviation of a Lipschitz function from its average value is extremely small.
Without entering into the details, in general $\mathcal{P}( |\,\mathcal{I}
- \mean{\mathcal{I}}\, | > t) \sim e^{-\alpha Nt^2}$ (with $\alpha$ a
constant), so that the deviation is of order $O(1/\sqrt{N})$. In other
words, almost all the vectors are ``typical'' with respect to Lipschitz
functions with the uniform measure.

The generalisation of formula \ref{eq:average_unoq} for an operator with
many different eigenvalues is quite complicated. The limit of small errors
has however a simple interpretation in terms of the variance of the
eigenvalue phases. For generic eigenvalues $\{e^{i\lambda_1}, \dots,
e^{i\lambda_N}\! \}$ of $\opR$, let
\begin{equation} \label{eq:defsigmalambda}
  \mean{\lambda} = \frac{\sum_j \lambda_j}{N} \qquad\mathrm{and}\qquad
  \sigma^2_\lambda = \frac{\sum_j (\lambda_j - \mean{\lambda})^2}{N}.
\end{equation}

The small errors limit corresponds to eigenvalue phases with small spread
($\sigma_\lambda \ll 1$); in this limit the trace $|\tr(\opR)|$ can be
approximated by retaining only the terms up to the second order,
\begin{displaymath}
  \Big| {\textstyle\sum_j} e^{i\lambda_j} \Big|^2 \!=
  \Big| {\textstyle\sum_j} e^{i[\lambda_j - \mean{\lambda}]} \Big|^2 \!\simeq
  N^2 \Big| 1 - \frac{\sigma_\lambda^2}{2} \Big|^2 \simeq
  N^2 (1 - \sigma^2_\lambda),
\end{displaymath}
so that formula \ref{eq:average2} gives a value of the integral
which depends only on the variance of the eigenvalue phases:
\begin{equation} \label{eq:average_small}
  \mathcal{I}_2(\opR) = 1 - A \sigma^2_\lambda + O(\sigma_\lambda^4).
\end{equation}

Problem \ref{eq:genaverage} can be generalised to involve more vectors with
correlated distributions, for instance a variable \ket{\psi} with uniform
distribution $d\mu(\psi)$ in a generic vector space, and another variable
\ket{\phi} with uniform distribution $d\mu(\phi|\psi)$ in the subspace
orthogonal to \ket{\psi}, like in \small
\begin{displaymath}
  \mathcal{J}\left( \!\!\begin{array}{r} \opR\apc{1}\\ *\apc{1} \end{array},
    \ldots, \begin{array}{r} \opR\apc{r}\\ *\apc{r} \end{array} \!\!\right) =
  \int d\mu(\psi)d\mu(\phi|\psi) \,\prod_{k=1}^r
  \bra{\psi}\opR\apc{k}\ket{\phi}_,^{*\apc{k}}
\end{displaymath}
\normalsize where $r$ is a number of operators and the symbol $*\apc{k}$
means either a null or a conjugation sign (i.e. some hermitian products can
be conjugated). In reference \cite{Brouwer97} it is shown that the number
of conjugations must be equal to that of ``non conjugations'' in order to
have a non-zero value for the integral. The $\mathcal{J}$ integrals can be
written as functions of the $\mathcal{I}$ integrals; as an example, the
case $r=2$ will be worked out here. The first step consists in inserting a
unitary operator $\opS_\psi$ mapping a fixed vector \ket{0} into
\ket{\psi}.  Changing the integration variable from $\ket{\phi}$ to
$\ket{\phi'} = \opS_\psi^\dagger\ket{\phi}$ allows to express the measure
in a simple way, because $d\mu(\phi')$ becomes simply the uniform measure
in the $(N-1)$-dimensional subspace $\mathrm{Span}\left\{\ket{i} \,|\, i
  \ne 0\right\}$: \small
\begin{eqnarray*}
  \lefteqn{\mathcal{J}_2(\opR) =
    \mathcal{J}\left( \!\!\begin{array}{r} \opR \\ {} \end{array},
      \begin{array}{r} \opR \\ * \end{array} \!\!\right) =
    \int d\mu(\psi)d\mu(\phi|\psi) |\bra{\psi}\opR\ket{\phi}|^2} \\
   &=& \int d\mu(\psi) \bra{\psi} \opR\opS_\psi \Big[ \int d\mu(\phi|\psi)
        \opS_\psi^\dagger\ket{\phi}\bra{\phi}\opS_\psi \Big] 
        \opS_\psi^\dagger\opR^\dagger \ket{\psi} \\
   &=& \int d\mu(\psi) \bra{\psi} \opR\opS_\psi \Big[ \int d\mu(\phi')
        \ket{\phi'}\bra{\phi'} \Big] \opS_\psi^\dagger\opR^\dagger \ket{\psi}.
\end{eqnarray*}
\normalsize

It follows from result \ref{eq:average1} that $\int d\mu(\phi') \ket{\phi'}
\bra{\phi'}$ can be replaced by $\frac{1}{N-1}\id_{N-1}$, where $\id_{N-1}$
is the identity in the appropriate subspace with $N-1$ dimensions. The
action of $\opS_\psi$ maps this identity to the projector onto the subspace
orthogonal to \ket{\psi}, that is $\opS_\psi\, \id_{N-1}\opS_\psi^\dagger =
\id - \ket{\psi}\bra{\psi}$, therefore
\begin{eqnarray*}
  \mathcal{J}_2(\opR)
   &=& \frac{1}{N-1} \int d\mu(\psi) \bra{\psi} \opR\opS_\psi\,
        \id_{N-1} \opS_\psi^\dagger\opR^\dagger \ket{\psi} \\
   &=& \frac{1}{N-1} \int d\mu(\psi) \bra{\psi} \opR \Big[
        \id - \ket{\psi} \bra{\psi} \Big] \opR^\dagger \ket{\psi}.
\end{eqnarray*}

Splitting the integral and inserting the result \ref{eq:average_small} for
$\mathcal{I}_2(\opR)$ shows once again that, in the limit of unitary $\opR$
close to the identity, the result depends only on the dimension of the
space and the spread of the eigenvalue phases $\sigma_\lambda$ :
\begin{equation} \label{eq:av_ain_m2}
  \mathcal{J}_2(\opR) = \frac{\mathcal{I}(\opR\opR^\dagger)
                       - \mathcal{I}_2(\opR)}{N-1}
    = \frac{N}{N^2-1} \,\sigma_\lambda^2 + O(\sigma_\lambda^4).
\end{equation}

\end{document}